\documentclass[aoas]{imsart}

\RequirePackage{amsthm,amsmath,amsfonts,amssymb}
\RequirePackage[authoryear]{natbib}
\RequirePackage[colorlinks,citecolor=blue,urlcolor=blue]{hyperref}
\RequirePackage{graphicx}

\usepackage{algorithm,algorithmic}
\usepackage{amsmath,amsfonts,amsthm,xcolor}
\usepackage{natbib}
\usepackage{graphicx}
\usepackage{subfig}
\usepackage{verbatim}

\usepackage{booktabs}
\usepackage{array}
\usepackage{multirow}
\usepackage{rotating}
\usepackage[normalem]{ulem}

\startlocaldefs

\newtheorem{theorem}{Theorem}

\newcommand{\red}[1]{#1}

\endlocaldefs

\begin{document}

\begin{frontmatter}
\title{Modeling panels of extremes}
\runtitle{Modeling panels of extremes}

\begin{aug}
\author[A]{\fnms{Debbie J.} \snm{Dupuis}\ead[label=e1]{debbie.dupuis@hec.ca}},
\author[B]{\fnms{Sebastian} \snm{Engelke}\ead[label=e2]{sebastian.engelke@unige.ch}}
\and
\author[C]{\fnms{Luca} \snm{Trapin}\ead[label=e3]{luca.trapin@unibo.it}}

\address[A]{Department of Decision Sciences,
			HEC Montr\'eal, 
			3000, chemin de la C\^ote-Sainte-Catherine,
			Montr\'eal (Qu\'ebec) H3T 2A7, Canada \printead{e1}}
                      \address[B]{Research Center for Statistics, University of Geneva, Boulevard du Pont d'Arve 40, 1205 Geneva, Switzerland, \printead{e2}}
\address[C]{Department of Statistics,
			University of Bologna,
			Via delle Belle Arti 41, 40126 Bologna, Italy \printead{e3}}

\end{aug}

\begin{abstract}
  Extreme value applications commonly employ regression techniques to capture cross-sectional heterogeneity or \red{time-variation} in the data. Estimation of the parameters of an extreme value regression model is notoriously challenging due to the small number of observations that are usually avai\-lable in applications. When repeated extreme measurements are collected on the same individuals, i.e., a panel of extremes is available, pooling the observations in groups can improve the statistical inference.
  We study three data sets related to risk assessment in finance, climate science, and hydrology.
  In all three cases, the problem can be formulated as an extreme value panel regression model with a latent group structure and group-specific parameters. We propose a new algorithm that jointly assigns the individuals to the latent groups and estimates the parameters of the regression model inside each group. Our method efficiently recovers the underlying group structure without prior information, and for the three data sets it provides improved return level estimates and helps answer important domain-specific questions.
\end{abstract}


\end{frontmatter}

\section{Introduction}
Extreme value theory provides the basis for the estimation of probabi\-lities and quantiles associated with extreme events \citep{coles2001introduction}. These quantities are used as inputs for managerial and policy decisions to prevent or limit the damage caused by extreme negative outcomes. Applications of extreme value theory commonly employ regression techniques to capture \red{distributional variations across individuals and over time} in datasets where covariates are available. For example, spatial applications model the parameters of the extreme value model as a function of geographical characteristics to explain heterogeneous beha\-vior across stations \citep{davison2012statistical} and climate applications typically include time as a covariate in the extreme value model to capture the effect of a changing climate \citep{katz2013statistical}. When repeated extreme measurements are available for several indivi\-duals, we may speak of a panel of extremes. \red{In particular, we will consider random variables $Y_{i,t}$ for individuals $i=1,\dots, N$ and time points $t=1,\dots, T$ that follow generalized extreme value distributions \citep{coles2001introduction} with parameters depending on a covariate vector $\mathbf{X}_{i,t} \in \mathbb{R}^K$ in a parametric way.}
In this case, practitioners would like to pool all observations to estimate the extreme value regression model. This idea of borrowing strength is very attractive in extreme value applications as the number of observations for the analysis is inherently small. It does present a practical challenge however as it requires observations to be homogeneous with respect to the model parameters.

The naive approach pools all available individuals into one group, regardless of the underlying structure. This may hide large differences in the model parameters and lead to misguided inference and sub-optimal quantile estimates. More often, empirical analyses rely on \textit{ad hoc} strategies to partition the individuals into homogeneous groups. 
For example, in spatial applications one may select groups of geographically close stations as homogeneous regions \citep{asadi2015extremes, overeem2008}; meteorological applications partition the data based on climatological similarity \citep{alila1999} or physiographical similarity and meteorological factors \citep{cheng1998};  while financial applications partition the data based on economic sectors, business lines \citep{mhalla2020extremal} or type of losses \citep{hambuckers2021smooth}. While this approach introduces flexibility in the extreme panel regression model, it requires \textit{a priori} domain know\-ledge.
The latter might not be available, or may be inconsistent with the data generating mechanism.

In this paper, we study three data sets where identification of the correct group structures is crucial for accurate statistical inference.
The first data set is for a financial study invol\-ving 48 assets where an extreme value regression with time-varying asset-specific covariates is used to tie financial losses to risk factors relevant to generate stress test scenarios. This is a critical tool to track firms' exposures to adverse extreme market events. Pooling across all assets
allows estimating the regression model with few observations per asset, but yields poor estimates of the relationship between financial losses and risk factors. Standard practice in finance would group assets based on the Standard Industrial Classification (SIC), but this might not fully explain the heterogeneity \citep{oh2020dynamic}.

The second study investigates the effect of climate change on extreme temperatures in the U.S.~Midwest. Data are for 127 locations where an extreme value regression \citep{zwiers1998, wang2016gev}
 models the impact of the global temperature anomaly (a common time-varying variable) on minimum temperatures, controlling for time-invariant location-specific covariates. The sought-after global mean/local extreme relation can be inferred at a single location, but short series yield large standard errors. An analysis based on homogeneous groups of locations could reduce standard errors. 
The third data set on flood risk assessment is based on 31 gauging stations in the Danube river basin where an extreme value regression with time-invariant station-specific characteristics is used to model spatial heterogeneity. Risk analysis based only on individual stations is insufficient since short record length would lead to large uncertainty of return level estimates. There are many methods to construct homogenous groups of stations \citep[e.g.,][]{mer2005, asa2018}, but many of them are heuristic or require domain knowledge. 

While the three data sets come from different areas and exhibit domain specific challenges, the general structure of the data can be cast as a panel regression problem for extremes. Throughout the paper, we assume that the individuals are partitioned according to an unknown latent group structure and that the extreme value regression model presents group-specific parameters. Our aim is to make joint inference on the group structure and the model parameters. This requires finding the correct number of groups and the right assignment of the individuals. A full-likelihood approach would require estimating the model parameters and the group assignment concurrently, but this problem is computationally infeasible. To solve this issue, we develop a novel expectation-maximization (EM) algorithm that is able to estimate the model parameters while uncovering the latent group structure in a data-driven way. In this perspective, we add to a growing literature on identifying latent group structures in panel data \citep{su2016identifying, gu2019panel, oh2020dynamic, wang2021identifying}. \red{We also contribute to recent literature on extreme value clustering methods for spatial data \citep{ Carreau2017, ReichShaby2018, RohrbeckTawn2021, vig2021}.}

Our new methodology offers a solution to the challenges encountered in the three data sets. 
Comparing our data-driven group structure with groupings based on domain knowledge, we find that our approach always yields superior predictions and offers more reliable inference.
In the first study, the SIC grouping only partially explains the 
heterogeneity in the assets, and our model with data-driven group assignments produces superior risk estimates.
In the second study, our grouped panel model yields more precise estimates, identifying locations where extreme temperatures are more severely impacted by the global temperature anomaly and crop yields are more threatened.
In the third study, our optimal panel model discovers coherent spatial
similarity without using domain knowledge, yields smaller BIC values
than grouping based on domain knowledge, and provides a good fit even
for stations with a lot of missing data.

The remainder of the paper is structured as follows.
In Section \ref{Sec:EVpanel}, after a brief introduction to extreme value theory, we present our panel model for extremes. Section \ref{Sec:Sim} studies the finite sample properties of the proposed EM algorithm. In Sections \ref{Sec:EA1} to \ref{Sec:EA3}, we use our panel model for extremes in our three studies to provide superior estimates and more reliable inference.
Section \ref{Sec:End} concludes. The Appendix contains additional results.

\section{A panel model for extremes}
\label{Sec:EVpanel}

\subsection{Extreme Value Theory}
Let $Z_1, \dots, Z_s$ be a sample of independent observations of a distribution $F$ and define the sample block maximum $Y^{(s)} = \max\{Z_1, \dots ,Z_s\}$, where $s$ is called the block size. The Fisher--Tippett--Gnedenko theorem \citep[e.g.,][Theorem 3.2.3]{embrechts1997modelling} states that if there exist sequences of normalizing constants $a_s$ and $b_s$ such that the normalized $Y^{(s)}$ converges in distribution to a non-degenerate limit distribution $G$,
\begin{align}\label{gev}
  \lim_{s\to \infty} \mathbb P\left(\frac{Y^{(s)} - a_s}{b_s} \leq y \right) = \red{G}(y) , 
\end{align}
then $G$ must be the \red{generalized extreme value} (GEV) distribution
\begin{align}\label{gev_form}
H(y \mid\mu, \sigma, \xi) = 
\begin{cases}
\exp\left\lbrace -\left(1+\xi \frac{y-\mu}{\sigma}\right)_+^{-1/\xi} \right\rbrace, &  \quad \xi \not= 0, \\
\exp\left\lbrace -\exp \left(-\frac{y-\mu}{\sigma}\right)\right\rbrace, & \quad \xi = 0,
\end{cases}
\quad y \in \mathbb R,
\end{align}
where $x_+= \max(0,x)$. In this case $F$ is said to be in the max-domain of attraction of the GEV distribution $H$. The parameters $\mu\in\mathbb R$, $\sigma>0$ and $\xi\in\mathbb R$ are the location, scale and shape parameters, respectively. The shape is the most important parameter since it characterizes the heaviness of the tail of the distribution $F$: if $\xi> 0$ then $F$ is heavy tailed (e.g., Cauchy, Pareto) and the GEV is a Fr\'echet distribution; if $\xi = 0$, then $F$ is light-tailed (e.g., Gaussian, exponential) and the GEV is a Gumbel distribution; if $\xi < 0$ then $F$ has a finite upper endpoint (e.g., uniform, beta) and the GEV is a Weibull distribution; see \cite{coles2001introduction} for details.

The limiting result~\eqref{gev} holds under a very mild assumption on the tail of $F$, which is satis\-fied for almost all relevant continuous distributions. Since the GEV distribution is the only possible limit for sample maxima, it is an asymptotically motivated model for the distribution of $Y^{(s)}$ for finite values of $s$. Suppose we have $N$ independent observations $Y^{(s)}_{1}, \dots, Y^{(s)}_{N}$ of the block maximum $Y^{(s)}$. We use maximum likelihood estimation to obtain the parameters of the GEV distribution that best approximate the distribution of~$Y^{(s)}$,
\begin{align}\label{gev_mle}
  (\hat \mu, \hat \sigma, \hat \xi) = \operatorname*{arg\,max}_{\mu, \sigma, \xi} \sum_{i=1}^N \log h (Y^{(s)}_{i} \mid \mu,\sigma,\xi),
\end{align}
where $\log h(\cdot \mid \mu,\sigma,\xi)$ is the log-likelihood of the GEV distribution
\begin{align}\label{gev_llh}
\log h (y\mid \mu,\sigma,\xi) = - \log(\sigma) - \left(1 + \frac{1}{\xi}\right) \log\left(1 + \xi \frac{y - \mu}{\sigma} \right) - \left(1 + \xi\frac{y - \mu}{\sigma} \right)^{-1/\xi}.
\end{align}
The maximum likelihood estimator is consistent and asymptotically normal if $\xi > -1/2$ \citep{smi1985, bue2017}. Under mild conditions on the dependence structure of a stationary time series, the GEV also emerges as the only possible non-degenerate limiting distribution for normalized maxima of blocks of observations from this series \citep{leadbetter1983extremes}. Asymptotic properties of the maximum likelihood estimator continue to hold in this setting \citep{bucher2018maximum}.

In applications, the GEV distribution is often fitted to observations of $Y^{(s)}$ to estimate return levels. The $p$-quantile of the GEV distribution is
\begin{align}\label{QGEV}
Q^{p}(\mu, \sigma, \xi) = 
\begin{cases}
\mu - \dfrac{\sigma}{\xi} \left[ 1 - \{-\log(p)\}^{-\xi} \right], &  \quad \xi \not= 0, \\
\mu - \sigma \log\{-\log(p)\}, & \quad \xi = 0,
\end{cases}
\end{align}
and the $S$th return level, for $S>1$, is defined as the $(1 - 1/S)$-quantile. It represents the level that is expected to be exceeded on average once in $S$ time periods, where the time period corresponds to the block length $s$. For instance, if $Y^{(s)}$ represents a yearly maximum, then $RL^S(\mu, \sigma, \xi)=Q^{(1 - \frac{1}{S})}(\mu, \sigma, \xi)$ is the $S$-year return level.

\subsection{A panel GEV regression model}
\label{Sec:SimplePanel}
 Panel data studies have flourished over the last years \citep{hsiao2007panel}. Some of the advantages of panel data compared to cross-sectional and time series data are: more efficient estimates of the model parameters; the possibility to test more complicated models; controlling the impact of omitted variables (fixed effects); generating more accurate predictions by pooling information across individuals. Nowadays, static and dynamic panel models are available for linear regression \citep{hsiao2014analysis}, count data regression \citep{cameron2015count}, discrete choice models \citep{greene2009discrete}, and volatility models \citep{pakel2011nuisance}. We consider a panel of maxima $Y_{i,t}$, with $i=1,\dots, N$ and $t=1,\dots, T$, extracted for $N$ individuals in $T$ blocks. We omit the superscript for the block size $s$, but still assume that $Y_{i,t}$ is a block maximum and can be reasonably approximated by a GEV distribution. We further let the GEV model parameters depend on a covariate vector $\mathbf{X}_{i,t} \in \mathbb{R}^K$ measured for each $i$th individual in each $t$th block. The panel GEV regression model is defined as
\begin{align}\label{gev_panel}
Y_{i,t}\mid \ \mathbf{X}_{i,t} \sim H(y\mid \mu_{i,t},\sigma_{i,t},\xi_{i,t}),
\end{align}
where the model parameters depend on the covariates as
\[
\begin{array}{rll}
\mu_{i,t} &=& \red{e_\mu} \left(\boldsymbol{\kappa}^{\top} \mathbf{X}_{i,t} \right), \\
\sigma_{i,t} &=& \red{e_\sigma} \left(\boldsymbol{\gamma}^{\top} \mathbf{X}_{i,t} \right), \\
\xi_{i,t} &=& \red{e_\xi} \left(\boldsymbol{\delta}^{\top} \mathbf{X}_{i,t} \right),
\end{array}
\]
where $\boldsymbol{\theta}=\left(\boldsymbol{\kappa},\boldsymbol{\gamma},\boldsymbol{\delta}\right) \in \boldsymbol{\Theta} \subset \mathbb{R}^P$ is a vector of regression parameters, and \red{$e_\mu$, $e_\sigma$, and $e_\xi$} are suitable link functions that can be adapted to the requirements in applications. As we are not interested in the cross-sectional dependence structure in $\mathbf{Y}_{t}=\left(Y_{1,t},\dots,Y_{N,t}\right)$, we can make inference on the marginals by estimating the model parameters by quasi maximum likelihood (QML), pooling information across individuals, i.e.,
\begin{equation} 
\widehat{\boldsymbol{\theta}}_{QML} = \operatorname*{arg\,max}_{\boldsymbol{\theta}} \sum_{i=1}^{N} \sum_{t=1}^{T} \log h (Y_{i,t}\mid \mu_{i,t},\sigma_{i,t},\xi_{i,t})
\label{gev_qml}
\end{equation} 
where the log-likelihood of the GEV distribution is defined in~\eqref{gev_llh}. Under the usual regularity conditions and as $T \rightarrow \infty$ \citep[see][]{chandler2007inference} we have
\[
\widehat{\boldsymbol{\theta}}_{QML} \xrightarrow{d} N(\boldsymbol{\theta}, \mathbf{H}^{-1}\mathbf{V}\mathbf{H}^{-1})
\]
where $\mathbf{H}$ is the expected Hessian of the log-likelihood and $\mathbf{V}$ is the covariance matrix of the score evaluated at the true parameter. Consistent estimates of these two quantities can be obtained as
\[
\widehat{\mathbf{H}} = \sum^T_{t=1}  \sum^N_{i=1} \left.\frac{\partial \mathbf{s}_{it} (\boldsymbol{\theta})}{\partial\boldsymbol{\theta}} \right|_{\widehat{\boldsymbol{\theta}}_{QML}},
\]
\[
\widehat{\mathbf{V}} = \sum^T_{t=1} \left( \sum^N_{i=1} \mathbf{s}_{it} (\widehat{\boldsymbol{\theta}}_{QML}) \right) \left( \sum_{i=1}^{N} \mathbf{s}_{it} (\widehat{\boldsymbol{\theta}}_{QML}) \right)^{\top},
\]
where $\mathbf{s}_{it} (\boldsymbol{\theta}) = \partial / \partial \boldsymbol{\theta} \log h (Y_{i,t}\mid \mu_{i,t},\sigma_{i,t},\xi_{i,t}) $.

\subsection{A grouped panel GEV regression model}
\label{Sec:GroupPanel}
To introduce flexibility in panel models, a fast-growing literature explores the existence of latent group structures: see \cite{su2016identifying} and \cite{wang2021identifying} for examples in linear and non-linear panel regression models; \cite{gu2019panel} for an example in panel quantile models; \cite{oh2020dynamic} for an example with dynamic copula models. We assume that each of the $N$ individuals is a member of one of $G\geq 1$ groups. We let $\boldsymbol{\tau} =\left(\tau_1,\dots,\tau_N\right)$ be the $N$-dimensional vector whose $i$th entry $\tau_i \in \{1,\dots,G\}$ denotes the group membership of the $i$th individual. The grouped panel GEV model is defined as
\begin{align}\label{grouped_panel}
Y_{i,t}\mid \mathbf{X}_{i,t} \sim H\left(y\mid \mu_{i,t}(\tau_i),\sigma_{i,t}(\tau_i),\xi_{i,t}(\tau_i)\right),
\end{align}
with
\begin{align}
\label{grouped_panel_parms}
\begin{array}{rll}
\mu_{i,t}(\tau_i) &=& \red{e_\mu} \left(\boldsymbol{\kappa}_{(\tau_i)}^{\top} \mathbf{X}_{i,t}\right) \\
\sigma_{i,t}(\tau_i) &=& \red{e_\sigma} \left(\boldsymbol{\gamma}_{(\tau_i)}^{\top} \mathbf{X}_{i,t}\right) \\
\xi_{i,t}(\tau_i) &=& \red{e_\xi} \left(\boldsymbol{\delta}_{(\tau_i)}^{\top} \mathbf{X}_{i,t} \right)
\end{array}
\end{align}
where $\boldsymbol{\theta}_{(g)}=\left(\boldsymbol{\kappa}_{(g)},\boldsymbol{\gamma}_{(g)},\boldsymbol{\delta}_{(g)}\right) \in \boldsymbol{\Theta} \subset \mathbb{R}^P$ is the vector of regression parameters for the $g$th group, $g \in \{1,\dots,G\}$. We denote the full parameter vector by $\boldsymbol{\theta} = \left( \boldsymbol{\theta}_{(1)}, \dots, \boldsymbol{\theta}_{(G)} \right)$.  In the existing literature, the individuals are typically grouped according to observed characteristics, such as physical distance in spatial applications and business lines in economic applications \citep{asadi2015extremes, mhalla2020extremal}. Given the group assignments, the parameters of the grouped panel GEV model can be estimated group-wise using the QML estimator in \eqref{gev_qml}. However, \textit{a priori} assignments may not provide the best fit to the data; see the applications in Sections \ref{Sec:EA1} and \ref{Sec:EA3}.

We therefore propose a data-driven method that jointly estimates the group assignments $\boldsymbol{\tau}$ and the vector of model parameters $\boldsymbol{\theta}$  based on an EM algorithm. \red{This algorithm iterates between estimating the regression parameters given the group assignments and estimating the group assignments given the regression parameters. Disentangling the estimation of the regression parameters and the group assignments drastically simplifies the estimation problem.}

Let $(Y_{i,t}, \mathbf{X}_{i,t})$ be the observations from the grouped panel GEV model, where $i=1,\dots, N$ and $t=1,\dots, T$. Our algorithm has the following steps.

\begin{enumerate}
	\item[0.] \textbf{Initialization}. For a fixed value of $G$, select an initial group assignment $\boldsymbol{ \widehat\tau}^{(0)}$ at random.
	\item[1.] \textbf{Iteration}.  We iterate between the following two steps until convergence. At the $j$th iteration, we have
	\begin{enumerate}
		\item[1.1] \textbf{Maximization}. Conditioning on the group structure $\boldsymbol{ \widehat \tau}^{(j-1)}$ identified at the previous iteration, we apply the QML estimator over each of the $G$ groups. The EM estimator is thus defined as $\widehat{\boldsymbol{\theta}}^{(j)} = \left( \widehat{\boldsymbol{\theta}}^{(j)}_{(1)}(\widehat{\boldsymbol{\tau}}^{(j-1)}), \dots, \widehat{\boldsymbol{\theta}}^{(j)}_{(G)}(\widehat{\boldsymbol{\tau}}^{(j-1)}) \right)$ with
		\[
		\widehat{\boldsymbol{\theta}}^{(j)}_{(g)}(\boldsymbol{\tau}) = \operatorname*{arg\,max}_{\boldsymbol{\theta}} \sum_{i:g=\tau_i} \sum_{t=1}^{T} \log h \left(Y_{i,t} \mid \mu_{i,t}(\tau_i),\sigma_{i,t}(\tau_i),\xi_{i,t}(\tau_i)\right), \quad g = 1,\dots, G.
		\]
		\item[1.2] \textbf{Expectation}. Given the estimated parameters $\widehat{\boldsymbol{\theta}}^{(j)}$, we can find the best group assignments $\widehat{\boldsymbol{\tau}}^{(j)}$ maximizing the individual contribution to the likelihood, i.e.
		\[
		\widehat{\tau}^{(j)}_i = \operatorname*{arg\,max}_{g\in\left\{1,\dots,G\right\}} \sum_{t=1}^{T} \log h (Y_{i,t} \mid \widehat{\mu}_{i,t}(g),\widehat{\sigma}_{i,t}(g),\widehat{\xi}_{i,t}(g)), \quad i =1,\dots, N,
		\]
		where $\widehat{\mu}_{i,t}(g)= \red{e_\mu} \left(\boldsymbol{\widehat{\kappa}}_{(g)}^{\top} \mathbf{X}_{i,t} \right)$, $\widehat{\sigma}_{i,t}(g)= \red{e_\sigma} \left(\boldsymbol{\widehat{\gamma}}_{(g)}^{\top} \mathbf{X}_{i,t}\right)$, $\widehat{\xi}_{i,t}(g) = \red{e_\xi} \left(\boldsymbol{\widehat{\delta}}_{(g)}^{\top} \mathbf{X}_{i,t}\right)$.
		This step only requires the comparison of $G$ constants for each individual.
	\end{enumerate}
      \end{enumerate}

      The output of the EM algorithm is the estimated group assignments $\widehat{\boldsymbol{\tau}}_T$ and the estimated group parameters $\widehat{\boldsymbol{\theta}}_T$, where we sometimes drop the subscript $T$ for the number of samples in time. \red{We call Step 1.2 an expectation step for familiarity even though we do not compute an expectation.  We rather use the alternative approach of assigning each observation to the likelier group \citep{McLachlanEM} as per \cite{oh2020dynamic}.}

      Theorem~\ref{thm1} in Appendix~\ref{app:consistency} provides a consistency result for this EM algorithm proving that for a fixed number of individuals $N$ and a growing number of samples per individual $T \to \infty$, the group assignments and estimated group parameters converge to their true counterparts, that is, we have the convergences in probability
      $$\widehat{\boldsymbol{\tau}}_T \xrightarrow{p} \boldsymbol{\tau}_0, \qquad \widehat{\boldsymbol{\theta}}_T \xrightarrow{p} \boldsymbol{\theta}_0, \qquad T \to \infty.$$
      \red{Our assumption on the sequence of samples $(Y_{i,t}, \mathbf{X}_{i,t})$, $i=1,\dots, N$, $t=1,\dots, T$, for growing $T$ is that they form a stationary and erdogic sequence. We note that this is fairly general since  despite the stationarity of the joint sequence, the conditional distributions $Y_{i,t}\mid \mathbf{X}_{i,t}$ vary across individuals and over time with the covariate vector $\mathbf{X}_{i,t}$ according to the panel model in~\eqref{grouped_panel}.}      
We also require a set of regularity conditions on the log-likelihood of the GEV distribution as a function of the model parameters $\boldsymbol{\theta}$. We note here that such regularity conditions are not always easy to verify for the GEV distribution because of the changing support, and, in general, it will depend on the link functions used in the grouped panel GEV model whether these conditions are met. The asymptotic theory of maximum likelihood estimation for the GEV under the most general conditions remains an active area of research even in the i.i.d.\ case, see \cite{dom2015}, \cite{bue2017} and \cite{dom2019} for recent results. 
      
Inference on the parameters of the grouped panel is performed as in the single group setting in Section \ref{Sec:SimplePanel}. Standard errors for the parameters of each group are computed assuming independence among observations in different groups.
An important topic in the clustering literature is the selection of the number of groups \citep{su2016identifying, oh2020dynamic}, and we face an analogous challenge. As the number of groups is unknown, we repeat our procedure for different values of $G$ and rely on the BIC to select the optimal number of groups $G^*$, i.e. 
\[
G^* = \operatorname*{arg\,min}_{G\in\mathbb{N}} \textrm{BIC}(G)
\]
where 
\[\textrm{BIC}(G) = - 2\sum_{g=1}^G \sum_{i:\widehat{\tau}_i=g} \sum_{t=1}^{T} \log h (Y_{i,t}\mid \widehat{\mu}_{i,t}(\widehat{\tau}_i),\widehat{\sigma}_{i,t}(\widehat{\tau}_i),\widehat{\xi}_{i,t}(\widehat{\tau}_i)) + \log(NT) \times PG,
\]
\red{and $P$ is the dimension of the parameter vector $\boldsymbol{\Theta}$ in~\eqref{grouped_panel} in each group.}

\section{Simulation study}
\label{Sec:Sim}
We design a comprehensive simulation study in order to assess the finite sample properties of our EM algorithm for the grouped panel GEV model in typical settings. We generate sample maxima $\mathbf{Y}_t = \left(Y_{1,t}, \dots, Y_{N,t} \right)$ for $t = 1,\dots,T$ according to the following model
\[
\begin{array}{l}
\mathbf U_t = \left(U_{1,t}, \dots, U_{N,t}\right) \sim \mathbf{C}_{\alpha},\\
\mathbf{Y}_t =  \left(H^{-1}_{1,t}(U_{1,t}), \dots, H^{-1}_{N,t}(U_{N,t}) \right),
\end{array} 
\]
where $\mathbf{C}_{\alpha}$ is a copula characterizing the cross-sectional dependence structure with dependence parameter $\alpha$, and $H_{i,t} (y) = H\left(y \mid \mu_{i,t}(\tau_i),\sigma_{i,t}(\tau_i),\xi_{i,t}(\tau_i)\right)$ is the marginal GEV distribution. We assume constant shape parameters through time and across individuals in the same group, i.e., $\xi_{i,t}(g)=\delta_{0,(g)}$, and time-varying location and scale parameters as functions of the vector of covariates $\mathbf{X}_{i,t}=\left(X^{(1)}_{i,t}, X^{(2)}_{i}\right)$. In particular,
\[
\begin{array}{l}
\mu_{i,t}(\tau_i) = \kappa_{0,(\tau_i)} + \kappa_{1,(\tau_i)} X^{(1)}_{i,t} + \kappa_{2,(\tau_i)} X^{(2)}_{i}, \\
\sigma_{i,t}(\tau_i) = \exp\left( \gamma_{0,(\tau_i)} + \gamma_{1,(\tau_i)} X^{(1)}_{i,t} + \gamma_{2,(\tau_i)} X^{(2)}_{i} \right),
\end{array}
\] 
with $\tau_i \in \left\{1,\dots,G \right\}$. We let $X^{(1)}_{i,t}$ evolve according to the factor model
\[
X^{(1)}_{i,t} = \omega + \lambda t + \beta f_t + \epsilon_{i,t},
\]
where $f_t\sim N(0,\nu_f)$ and $\epsilon_{i,t}\sim N(0,\nu_i)$. These dynamics are designed to characterize a time-varying individual-specific covariate as in the financial risk application of Section \ref{Sec:EA1}. We let $X^{(2)}_i$ be uniformly distributed within the interval $\left(\underline{u}, \overline{u}\right)$ to characterize a time-invariant individual-specific covariate like the spatial characteristics used in the climatological and hydrological applications of Sections \ref{Sec:EA2} and \ref{Sec:EA3}, respectively.

For $100$ repetitions, we generate samples $\left\{Y_{i,t}, \mathbf{X}_{i,t} \right\}^{N,T}_{i=1,t=1}$ with $N=24$ and $T \in \left\{10,20,50\right\}$. The performance of the EM algorithm should improve as $T$ increases. We consider three copula functions to assess how the algorithm behaves under different dependence structures: the independence copula ($\red{\mathbf{C}}^{Ind}$) imposing zero cross-sectional dependence; the Gaussian copula ($\red{\mathbf{C}}^{Gauss}_{0.5}$) with constant correlation coefficient $\alpha=0.5$ across the individuals, implying a moderate cross-sectional dependence ($\text{Kendall's} \; \tau=0.33$) but zero tail dependence; the Gumbel copula ($\red{\mathbf{C}}^{Gum}_{2}$) with parameter $\alpha=2$ implying similar moderate cross-section dependence ($\text{Kendall's} \; \tau=0.5$), but positive tail dependence. We assume that the true number of groups is $G_0=4$ and fix the model parameters as in Table~\ref{Tab:ParSim}. We assign an equal number of individuals $N/G_0=6$ to each group. We estimate the grouped panel GEV model considering $G\in\left\{1,\dots,6\right\}$ and select the optimal number of groups using the BIC. Table \ref{Tab:ClusteringSimu} reports the performance of this approach. The ability of the BIC to recover the correct number of groups is already very good with only 20 time points and is perfect when $T=50$. The quality of the assignment when $G=4$, as measured by the Rand index \citep{rand1971objective}, a measure of similarity between two partitions, is good for small $T$ and it is almost perfect for $T=50$. An interesting aspect emerging from Table \ref{Tab:ClusteringSimu} is that the ability of the BIC to recover the correct number of groups grows with increasing cross-sectional dependence. Stronger dependence helps the group identification because it reduces the variance among the individuals in the same group. Consequently, as this variance is lower, the variance of the estimator must increase, as expected by QML estimation under dependence.

\begin{figure}[htbp]
	\centering
	\includegraphics[width=\linewidth]{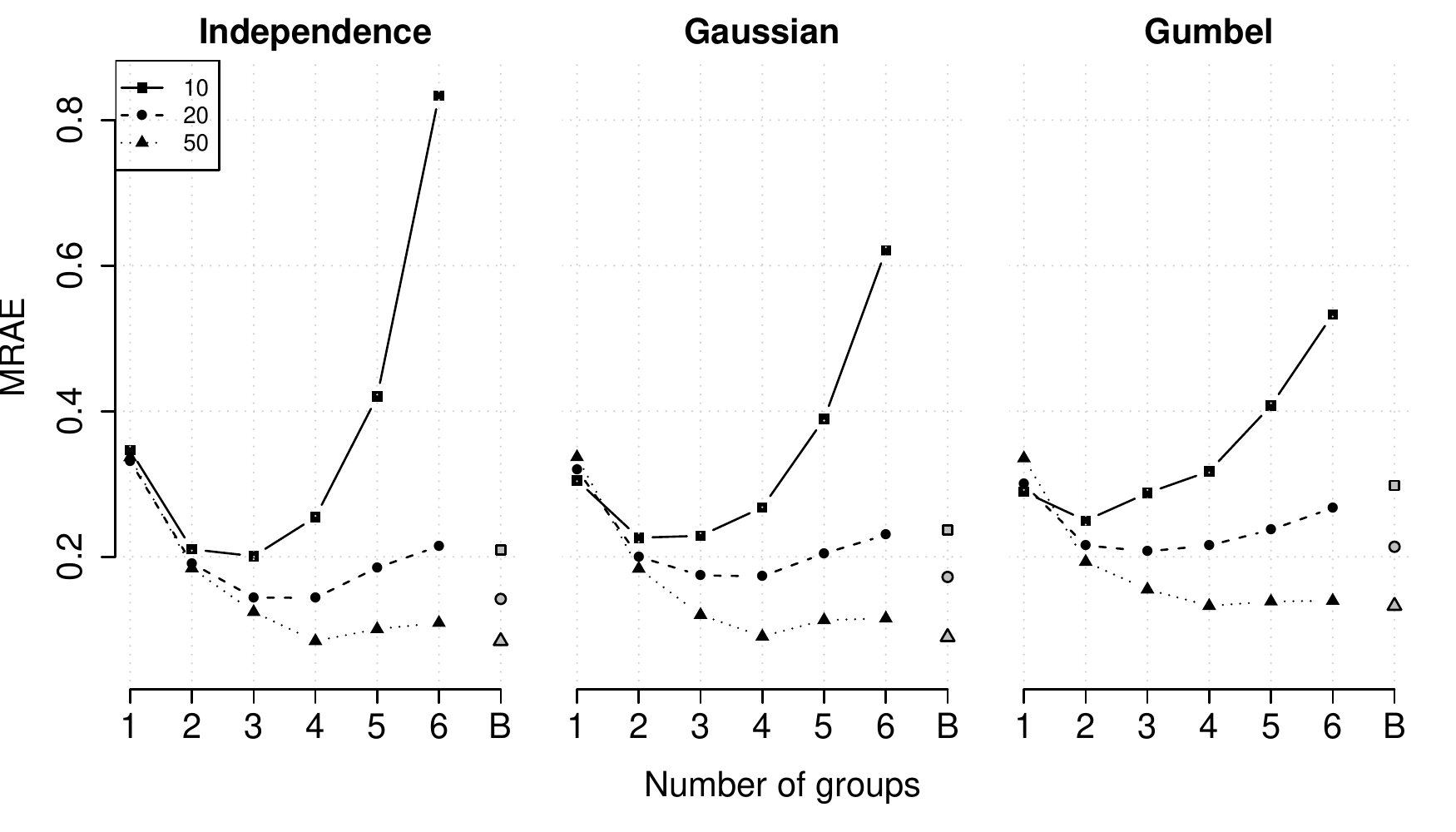}
	\caption{Median MRAE when estimating $Q^{0.99}_{it}$ using the model with $G\in\left\{1,\dots,6\right\}$ and the BIC selected model (B) over 100 replications: independence ($\mathbf{C}^{Ind}$), Gaussian ($\mathbf{C}^{Gauss}_{0.5}$) and Gumbel ($\mathbf{C}^{Gum}_2$) dependence.}
	\label{Fig:QuantileMAE}
\end{figure}

As applications are often concerned with the estimation of high quantiles of the GEV distribution, we also assess the performance of the EM algorithm from this perspective. First, we assess the quality of the quantile estimates for the different group structures obtained with the algorithm. We compute the mean relative  absolute error (MRAE) between the true 99th quantiles $Q^{0.99}_{it}$ (see equation~\eqref{QGEV}) and those estimated with $G\in\left\{1,\dots,6\right\}$. Figure~\ref{Fig:QuantileMAE} shows that the quality of the estimates is poor when the panel size is very small ($T=10$). As $T$ increases, $G=G_0=4$ provides the best quantile estimates. Interestingly, even though stronger dependence improves the group assignments (Table~\ref{Tab:ClusteringSimu}), the quality of the quantile estimates deteriorates as the dependence increases. This is coherent with the higher uncertainty of the QML estimator under dependence. Second, we assess the quality of the quantile estimates under the BIC-selected number of groups. We compute the MRAE between $Q^{0.99}_{it}$ and the quantiles estimated with the selected model. Figure~\ref{Fig:QuantileMAE} shows that the quantiles estimated with the BIC-selected model attain almost optimal performance in terms of MRAE compared to the pre-selected number of groups for $T$ as small as 20.

\begin{table}[htbp]
	\centering
	\caption{True value of parameters used in simulations to study the finite sample properties of the EM algorithm.}
	\begin{tabular}{ccccccrl}
		\toprule
	   & \multicolumn{4}{c}{Group parameters}    & \multicolumn{3}{c}{Covariate parameters} \\
		\midrule
		                 & $g=1$ & $g=2$ & $g=3$ & $g=4$ & & \\
		\cmidrule{2-5}       
		$\kappa_{0,(g)}$ & 3.10  & 3.40  & 3.20  & 3.10  & &  $\;\;\;\;\;\;\omega$ & -0.8 \\
		$\kappa_{1,(g)}$ & 2.40  & 1.40  & 1.10  & 1.70  & &  $\lambda$ & 0.4/T \\
		$\kappa_{2,(g)}$ & 2.00  & 1.00  & 0.50  & 1.50  & &  $\beta$  & 0.8   \\
		$\gamma_{0,(g)}$ & -0.05 &-0.15  & -0.20 &-0.10  & &  $\nu_f$  & 0.5   \\
		$\gamma_{1,(g)}$ & 0.10  & 0.06  & 0.04  & 0.08  & &  $\nu_I$	& 0.5   \\
		$\gamma_{2,(g)}$ & 0.17  & 0.07  & 0.02  & 0.12  & &  $\underline{u}$ & 2 \\
		$\delta_{0,(g)}$ & 0.30  & 0.27  & 0.24  & 0.20  & &  $\overline{u}$  & 6 \\
		\bottomrule
	\end{tabular}
	\label{Tab:ParSim}
\end{table}

\begin{table}
	\centering
	\caption{Performance of the BIC in selecting the correct number of groups and average Rand index computed between the true and estimated group structures when $G_0=4$ over $100$ replications:  independence ($\mathbf{C}^{Ind}$), Gaussian ($\mathbf{C}^{Gauss}_{0.5}$) and Gumbel ($\mathbf{C}^{Gum}_2$) dependence.}
	\begin{tabular}{ccccccc}
		\toprule
		Size & \multicolumn{2}{c}{$\mathbf{C}^{Ind}$} & \multicolumn{2}{c}{$\mathbf{C}^{Gauss}_{0.5}$} & \multicolumn{2}{c}{\textit{$\mathbf{C}^{Gum}_2$}} \\
		\midrule
		& BIC & Rand & BIC & Rand & BIC & Rand \\ 
		\cmidrule(l){2-3} \cmidrule(l){4-5} \cmidrule(l){6-7}
		$T=10$ & 24\% & 88\%  & 34\% & 91\% & 42\% & 93\% \\
		$T=20$ & 78\% & 94\%  & 92\% & 97\% & 88\% & 98\% \\
		$T=50$ &100\% & 99\%  & 99\% & 99\% &100\% & 99\% \\
		\bottomrule
	\end{tabular}
	\label{Tab:ClusteringSimu}
\end{table}

\pagebreak

\section{Financial risk management}
\label{Sec:EA1}
Extreme value theory offers suitable instruments for stress testing of extreme losses \citep{bis2001}. The GEV distribution in~\eqref{gev_form} is commonly employed in the financial literature to model maximum losses over a fixed horizon, both in unconditional static \citep{bali2003extreme, mcneil2015quantitative} and conditional dynamic \citep{zhao2018modeling} settings. In a stress testing setup, a GEV regression model ties maximum losses to a time-varying market risk factor to study how changes to the latter affect large losses. An example for such a risk factor is the semi-variance, \red{the average of the squared deviations of values that are less than the mean. It is a measure for downside risk, unlike the variance which provides a measure of volatility.  The semi-variance is} a widely used measure of downside risk in finance \citep{barn2010}, which is strongly related to macroeconomic uncertainty \citep{jurado2015measuring,segal2015good}.

We consider an institutional investor diversifying across 48 U.S. industry portfolios and interested in a stress test analysis of the portfolios' annual maximum daily loss using the portfolios' annual realized semi-variance as a risk factor.  
Table \ref{Tab:Industries} shows the 48 U.S. industry portfolios for which we collect daily returns from 1950 to 2018\footnote{Details on the composition of the portfolios and data are available from the Kenneth French Data Library at \url{http://mba.tuck.dartmouth.edu/pages/faculty/ken.french/data_library.html}.}. Let $Z_{i,t,j}$ be the daily return of the $i$th industry portfolio on the $j$th day of the $t$th year, with $i = 1,\dots, 48$ and $t = 1,\dots, 69$. We define the annual maximum loss $Y_{i,t} = \max_{j=1,\dots, s} \left(-Z_{i,t,j}\right)$ and the annual realized semi-variance $RS_{i,t} = \sum^{s}_{j=1} Z^2_{i,t,j} \mathbb{I}{\{Z_{i,t,j} < 0\}}$, where $\mathbb{I}{\{\cdot\}}$ denotes the indicator function.
Accurate stress testing requires a model that properly characterizes the variation in the extreme quantiles of such industry portfolios to changes in the annual realized semi-variance.
We model the panel of maximum losses with the following grouped panel GEV model
\[
\begin{array}{rll}
\log \left(\mu_{i,t}\left(\tau_i\right)\right) &=& \kappa_{0,(\tau_i)} + \kappa_{1,(\tau_i)} \log \left(RS_{i,t}\right) \\
\log \left(\sigma_{i,t}\left(\tau_i\right)\right) &=& \gamma_{0,(\tau_i)} \\
\log \left(\xi_{i,t}\left(\tau_i\right)\right) &=& \delta_{0,(\tau_i)} + \delta_{1,(\tau_i)} \log \left(RS_{i,t}\right)
\end{array}
\]
with $\tau_i\in \left\{1,\dots,G\right\}$. The realized semi-variance in the location parameter accounts for the heteroskedasticity characterizing financial data. As periods of high volatility should lead to larger extremes, we expect the $\kappa_{1}$ parameters to be positive. Similarly, we model the shape parameter $\xi_{i,t}$ as a function of the realized semi-variance to account for the changing nature of tail risk. Previous analyses have shown that the shape parameter changes over time and it is positively associated to market volatility and economic uncertainty \citep{zhao2018modeling,massacci2017tail}, therefore we expect the $\delta_{1}$ parameters to be positive. We consider the logarithm of $RS_{i,t}$ as it is convenient from a modeling perspective \citep{bee2019realized}.

We first estimate the model using all the available observations simultaneously, i.e., setting $G=1$. Table \ref{Tab:MacroEst} shows that the QML estimate for $\kappa_{1,(g)}$ is positive and strongly statistically significant.
The parameter $\delta_{1,(g)}$ is positive but not statistically significant. This would lead us to conclude that the variation in the realized semi-variance is not informative of tail risk in the stock market, as measured by the shape parameters $\xi_{i,t}$. However, a more careful analysis shows that the omitted heterogeneity is leading to misguided inference. We explore possible group structures and estimate the panel GEV model with the EM algorithm using $G \in \left\{2,3,4,5,6\right\}$. The BIC-based optimal number of groups is $G^*=4$. Table \ref{Tab:Industries} shows the optimal group assignments $\boldsymbol{\widehat{\tau}}^*$ and Table \ref{Tab:MacroEst} shows the estimated parameters. The results suggest that there is strong group heterogeneity in the panel, particularly on the constant terms of the regression, i.e., the parameters $\kappa_{0}$, $\gamma_{0}$, $\delta_{0}$. Estimates for $\kappa_{1,(g)}$ are still positive and strongly statistically significant for each $g\in\left\{1,\dots,G\right\}$. The size of the $\delta_{1}$ parameters is now larger and statistically significant at the $5\%$ level for two of the four groups. The benefit of the grouped panel GEV model from a stress testing perspective is clear when comparing the 90th $(Q^{0.90}_{it})$ and 95th $(Q^{0.95}_{it})$ quantiles computed by \eqref{QGEV} for $G=1$ and for the optimal group structure $\boldsymbol{\widehat{\tau}}^*$. Table \ref{Tab:Vio} reports summary statistics for the quantile exceedances on each industry portfolio, i.e., $V^p_{i}=\frac{1}{T} \sum_{t=1}^T \mathbb I \left\{Y_{it}>Q^p_{it}\right\}$, with $p\in\left\{0.90, 0.95\right\}$.
While the median number of exceedances across portfolios is similar when using one group or the optimal assignments $\boldsymbol{\widehat{\tau}}^*$, the spread is considerably larger in the case of the former, both for the 90th and 95th quantiles, highlighting the greater accuracy of the latent group panel GEV model.

Standard practice in finance would group industry portfolios based on the Standard Industrial Classification (SIC) \citep{oh2020dynamic}, and for comparison we carry out our previous calculations based on this group assignment $\boldsymbol\tau^{SIC}$. Table \ref{Tab:Industries} shows the group assignments and Table \ref{Tab:MacroEst} shows the estimated parameters. Estimates uncover the heterogeneity in the $\delta_{1}$ parameters associated to the realized semi-variance, thus improving upon the $G=1$ estimates, but the group assignments are quite different from those identified with the EM algorithm and the corresponding maximized log-likelihood value is much smaller. 
As for quantile exceedances on each industry portfolio,
Table \ref{Tab:Vio} shows that the medians are similar, but the spread is greater when using $\boldsymbol\tau^{SIC}$ rather than $\boldsymbol{\widehat{\tau}}^*$, particularly on the 90th quantile. This suggests that the SIC grouping only partially explains the heterogeneity in the panel GEV model and that a data-driven procedure to group the portfolios is beneficial. 

\begin{table}[htbp]
	\caption{Industry portfolios and corresponding group assignments obtained by the EM algorithm $(\boldsymbol{\widehat{\tau}}^*)$ and defined by the SIC $(\boldsymbol{\tau}^{SIC})$.}
	\scriptsize
	\begin{tabular}{clllclll}
		\toprule
		\textit{Name} & \textit{Description} & $\boldsymbol{\widehat{\tau}}^*$ & $\boldsymbol{\tau}^{SIC}$ &  \textit{Name} & \textit{Description} & $\boldsymbol{\widehat{\tau}}^*$ & $\boldsymbol{\tau}^{SIC}$  \\
		\midrule
		Agric & Agriculture    & 4  & 1 & Ships&  Shipbuilding, Railroad Equipment & 4 & 2\\
		Food  & Food Products  & 2  & 1 &  Guns &  Defense & 1 & 2\\
		Soda  & Candy \& Soda  & 4  & 1 &  Gold &  Precious Metals & 4 & 5\\  
		Beer  & Beer \& Liquor & 1  & 1 &  Mines&  Non-Metallic and Industrial Metal Mining & 4 & 5 \\
		Smoke & Tobacco Products  & 2  & 1  &  Coal &  Coal & 4  & 2\\
		Toys  & Recreation        & 1  & 1  &  Oil  &  Petroleum and Natural Gas & 4 & 2 \\
		Fun   & Entertainment     & 1  & 5  &  Util &  Utilities & 2  & 2\\
		Books & Printing and Publishing & 3 & 1    &  Telcm&  Communication & 3 & 3  \\
		Hshld & Consumer Goods          & 3 & 1   &  PerSv&  Personal Services & 1 & 1\\
		Clths& Apparel                  & 3 & 1  &  BusSv&  Business Services & 3 & 3 \\
		Hlth & Healthcare               & 4 & 4  &  Comps&  Computers & 4 & 3\\
		MedEq& Medical Equipment   & 3      & 4  &  Chips&  Electronic Equipment & 4 & 3 \\
		Drugs& Pharmaceutical Products & 4 & 4 & LabEq&  Measuring and Control Equipment & 4 & 3 \\
		Chems& Chemicals     & 1        &2    & Paper&  Business Supplies & 1 & 2\\
		Rubbr& Rubber and Plastic Products & 1 & 2 & Boxes&  Shipping Containers & 1 & 2 \\
		Txtls& Textiles      & 3       &  1      &  Trans&  Transportation & 3 & 5\\
		BldMt& Construction Materials & 3 & 5    &  Whlsl&  Wholesale &2  & 1\\
		Cnstr& Construction         & 1   & 5    &  Rtail&  Retail &3 & 1\\ 
		Steel& Steel Works         & 4    & 5    &  Meals&  Restaraunts, Hotels, Motels & 3 & 5 \\
		FabPr& Fabricated Products & 4    & 2   &  Banks&  Banking & 2 & 5\\
		Mach & Machinery           & 3    & 2    &  Insur&  Insurance & 3 & 5\\
		ElcEq& Electrical Equipment & 1   & 2    &  RlEst&  Real Estate & 1 & 5 \\
		Autos& Automobiles and Trucks & 1 & 2    &  Fin  &  Trading & 3 & 5\\
		Aero & Aircraft               & 1 & 2   &  Other&  - & 1 & 5 \\
		\bottomrule
	\end{tabular}
	\label{Tab:Industries}
\end{table}

\begin{table}[htbp]
	\centering
	\caption{Parameter estimates and log-likelihood values ($LLH$) of the models for the industry portfolios: single group $G=1$, optimal group structure $\boldsymbol{ \widehat \tau}^*$ from the EM algorithm, and SIC-based $\boldsymbol{\tau}^{SIC}$. Standard errors in parentheses.}
	\begin{tabular}{ccccccccccc}
		\toprule
	    & $G=1$ & \multicolumn{4}{c}{$\boldsymbol{ \widehat \tau}^*$} & \multicolumn{5}{c}{$\boldsymbol{\tau}^{SIC}$}    \\
		\midrule
		&                    & $g=1$ & $g=2$ & $g=3$ & $g=4$  & $g=1$ & $g=2$ & $g=3$ & $g=4$ & $g=5$\\
		\cmidrule(l){3-6} \cmidrule(l){7-11}       
		$\kappa_{0,(g)}$     & -0.88  & -1.23  & -1.12  & -0.98  & -0.89  & -0.91 & -0.89 & -0.74 & -0.82 & -0.91 \\
		                     & (0.07) & (0.05) & (0.16) & (0.04) & (0.05) & (0.09) & (0.07) & (0.39) & (0.25) & (0.06)  \\
		$\kappa_{1,(g)}$     & 0.43   & 0.49   & 0.49   & 0.45   & 0.43  & 0.43 &  0.43 &  0.40 &  0.42 &  0.43  \\
		                     & (0.02) & (0.01) & (0.04) & (0.01) & (0.01) & (0.02) & (0.02) & (0.09) & (0.06) & (0.01) \\
		$\gamma_{0,(g)}$     & -0.37  & -0.47  & -0.73  & -0.57  & -0.21  & -0.50 & -0.46 & -0.30 & -0.32 & -0.42 \\
		                     & (0.08) & (0.07) & (0.19) & (0.09) & (0.08) & (0.08) & (0.11) & (0.19) & (0.15) & (0.08) \\
		$\delta_{0,(g)}$     & -3.63  & -6.03  & -5.83  & -4.89  & -5.61 & -5.11 & -3.54 & -3.62 & -6.53 & -5.09 \\
		                     & (1.36) & (2.25) & (3.73) & (2.72) & (2.18) & (1.36) & (1.11) & (4.46) & (5.69) & (1.81)  \\
		$\delta_{1,(g)}$     &  0.33  &  0.81  &  0.84  &  0.68  &  0.65 & 0.71 & 0.41 & 0.30 & 0.87 &  0.60 \\
		                     & (0.22) & (0.42) & (0.52) & (0.55) & (0.33) & (0.27) & (0.23) & (0.78) & (0.91) & (0.29) \\
		\cmidrule(l){2-2} \cmidrule(l){3-6} \cmidrule(l){7-11}   
		$LLH$ & -4225.01 & \multicolumn{4}{c}{-4065.24} & \multicolumn{5}{c}{-4199.68} \\
		\bottomrule
	\end{tabular}
	\label{Tab:MacroEst}
\end{table}

\begin{table}[htbp]
	\centering
	\caption{Summary statistics for the average number of exceedances of the 90th and 95th quantile for 48 U.S. industry portfolios: minimum ($Min$), 25th quantile ($\red{Q_1}$), median ($Med$), 75th quantile ($\red{Q_3}$), and maximum ($Max$). The expected number of exceedances is 0.10 and 0.05 for the 90th and 95th quantile, respectively.}
	\begin{tabular}{ccccccccccc}
		\toprule
		& \multicolumn{5}{c}{90$th$ quantile} & \multicolumn{5}{c}{95$th$ quantile}  \\
		\midrule
		& $Min$ & $Q_1$ & $Med$ & $Q_3$ & $Max$ & $Min$ & $Q_1$ & $Med$ & $Q_3$ & $Max$  \\
		\cmidrule(l){2-6} \cmidrule(l){7-11}
		$G=1$   & 0.057 & 0.098 & 0.116 & 0.134 & 0.261 & 0.014 & 0.041 & 0.058 & 0.087 & 0.161 \\
		$\boldsymbol{\widehat{\tau}}^*$ & 0.072 & 0.101 & 0.116 & 0.130 & 0.203 & 0.018 & 0.043 & 0.058 & 0.072 & 0.125 \\
		$\boldsymbol{\tau}^{SIC}$ & 0.072  & 0.089 & 0.116 & 0.130 & 0.247 & 0.018 & 0.043 & 0.058 & 0.072 & 0.161 \\ 
		\bottomrule
	\end{tabular}
	\label{Tab:Vio}
\end{table}

\newpage
\section{Effect of climate change on extreme temperatures}
\label{Sec:EA2}

\begin{figure}[htbp]
\begin{center}
        \null{\vspace{-1.2in}}
\includegraphics[width=\linewidth]{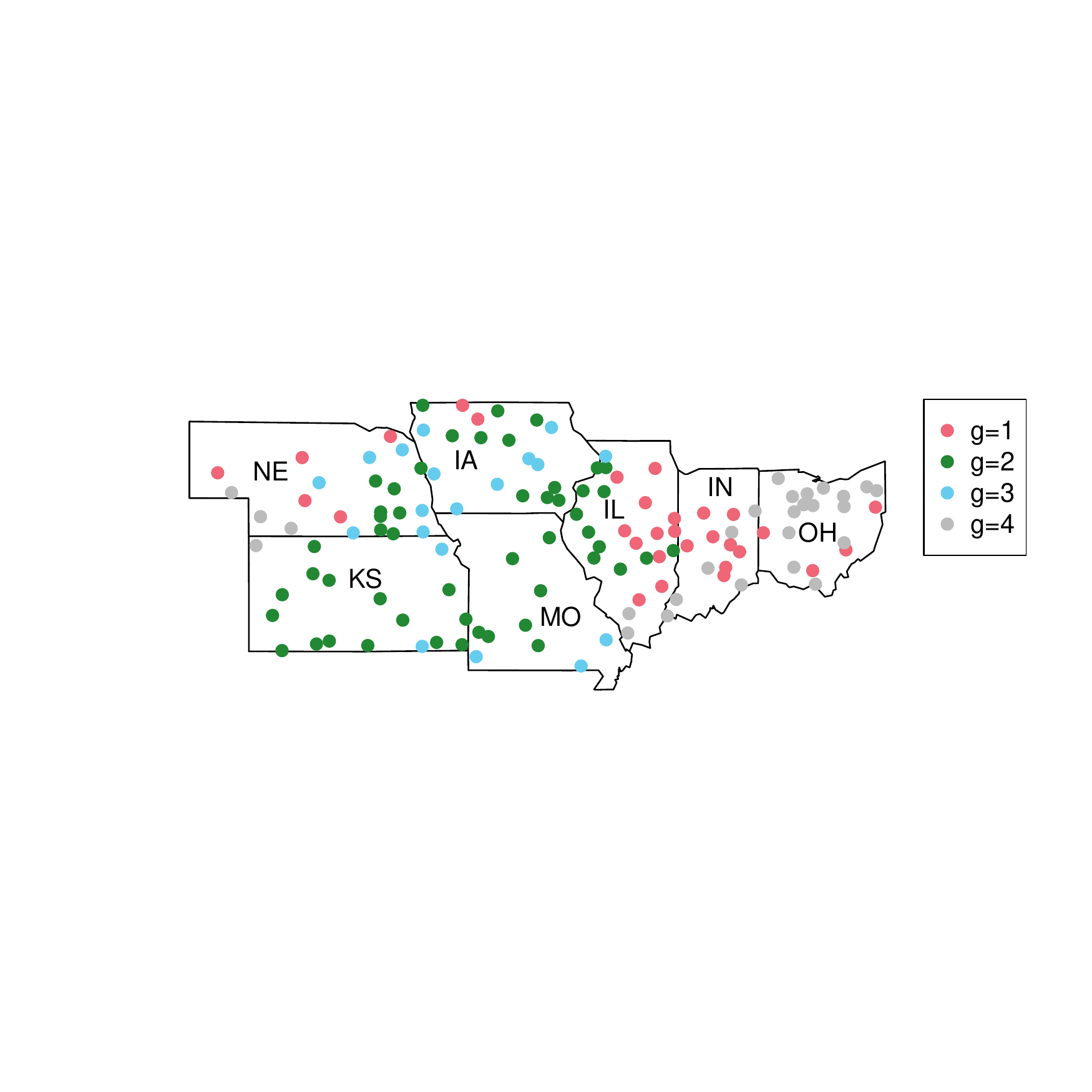}
        \null{\vspace{-2.1in}}
\end{center}
        \caption{Seven U.S.\ Midwest states and 127 weather stations. 
        The $G^*=4$ different colors of the stations correspond to the
        optimal grouping for the panel GEV model fitted to 
        the negative annual daily minimum temperature.}
        \label{Fig:MinTempBIC}
\end{figure}

There is a large literature on global mean temperature change and an increasing literature focusing on trends of tempera\-ture extremes, see respectively e.g. \cite{hansen2010} and \cite{papalexiou2018}, and references therein. The global mean tempe\-rature has recently been increasing at an increasing rate, but local temperature extremes have not always changed at the same rate, or even in the same direction. Temperature extremes have well-documented detrimental health and 
social impacts \citep{ipcc}. Nighttime warming reduces crop yields \citep{garcia2015, sadok2020} and can have large economic consequences for crop-producing regions. Assessing the global mean to local extreme temperature relationship makes the effects of climate change more relatable to economically vulnerable constituents and can guide public policy at the local level. Extreme value theory and the latent group panel GEV model developed in Section \ref{Sec:GroupPanel} allow us to seek this relationship 
for the crop-producing region in the U.S.\ Midwest in Figure \ref{Fig:MinTempBIC}.

The GEV distribution in~\eqref{gev_form} is a widely used model for
annual daily minimum tempe\-ratures \citep{zwiers1998, wang2016gev}.
Including the global land temperature anomaly as a covariate in
the GEV location and scale parameters allows us to infer the 
sought-after global mean/local extreme
relation at a given location, but short series yield large
standard errors and a panel model could provide more precise estimates.
We consider the negative annual daily minimum temperature $Y_{i,t}$ 
(in $\null^{\circ}$C),  $i=1,\dots,N$, $t=1,\dots,T$, at $N=127$ stations
in seven crop-producing states in the U.S.\ Midwest (Ohio, Indiana,
Illinois, Iowa, Missouri, Nebraska, and Kansas) during $T=99$ years;
see Figure \ref{Fig:MinTempBIC} for the locations of the weather
stations.
Data are available for
the 1912 to 2010 period in the {\tt USHCNTemp} dataset of the R package
{\tt SpatialExtremes} \citep{spatialextremes}.
Elevation and latitude should explain some of the variation in extremes
across the U.S.\ Midwest,
so we let the parameters of the panel GEV model to vary 
as a function of 
$\mathbf{X}_{i,t} = \left(\text{elev}_i, \text{lat}_i, \text{anom}_t \right)$
with
\begin{equation}
\begin{array}{rll}
\mu_{i,t}(\tau_i) &=& \kappa_{0,(\tau_i)} + \kappa_{1,(\tau_i)} \text{elev}_i +
              \kappa_{2,(\tau_i)} \text{lat}_i + \kappa_{3,(\tau_i)} \text{anom}_t \\
\log(\sigma_{i,t}(\tau_i)) &=& \gamma_{0,(\tau_i)} + \gamma_{1,(\tau_i)} \text{elev}_i +
              \gamma_{2,(\tau_i)} \text{lat}_i + \gamma_{3,(\tau_i)} \text{anom}_t \\
\xi_{i,t}(\tau_i) &=& \delta_{0,(\tau_i)}
\end{array}
\label{eq:TempPanelGG}
\end{equation}
for $\tau_i \in \left\{1,\dots,G\right\}$, where $\text{elev}_i$ and $\text{lat}_i$ denote, respectively, the elevation (in \red{$10^{3}$}~feet) and (normalized) latitude at station $i$, and $\text{anom}_t$ denotes the annual global land anomaly (in $\null^{\circ}$C) in year $t$.  Elevation and latitude are provided in the {\tt USHCNTemp} dataset. Annual global land anomalies are available at  \url{https://www.ncdc.noaa.gov/monitoring-references/faq/anomalies.php#anomalies}. We expect $\kappa_{0,(g)}$ to capture some of the group-specific idiosyncrasies, and $\kappa_{1,(g)}$ and $\kappa_{2,(g)}$ to be positive. The tail index $\delta_{0,(g)}$ should be negative as annual daily minimum temperature has a lower bound. The effect of the global anomaly on the location and scale parameters of the GEV, as measured by $\kappa_{3,(g)}$ and $\gamma_{3,(g)}$, respectively, is of interest.

\begin{figure}[htbp]
  \centering
        \includegraphics[width=0.49\linewidth]{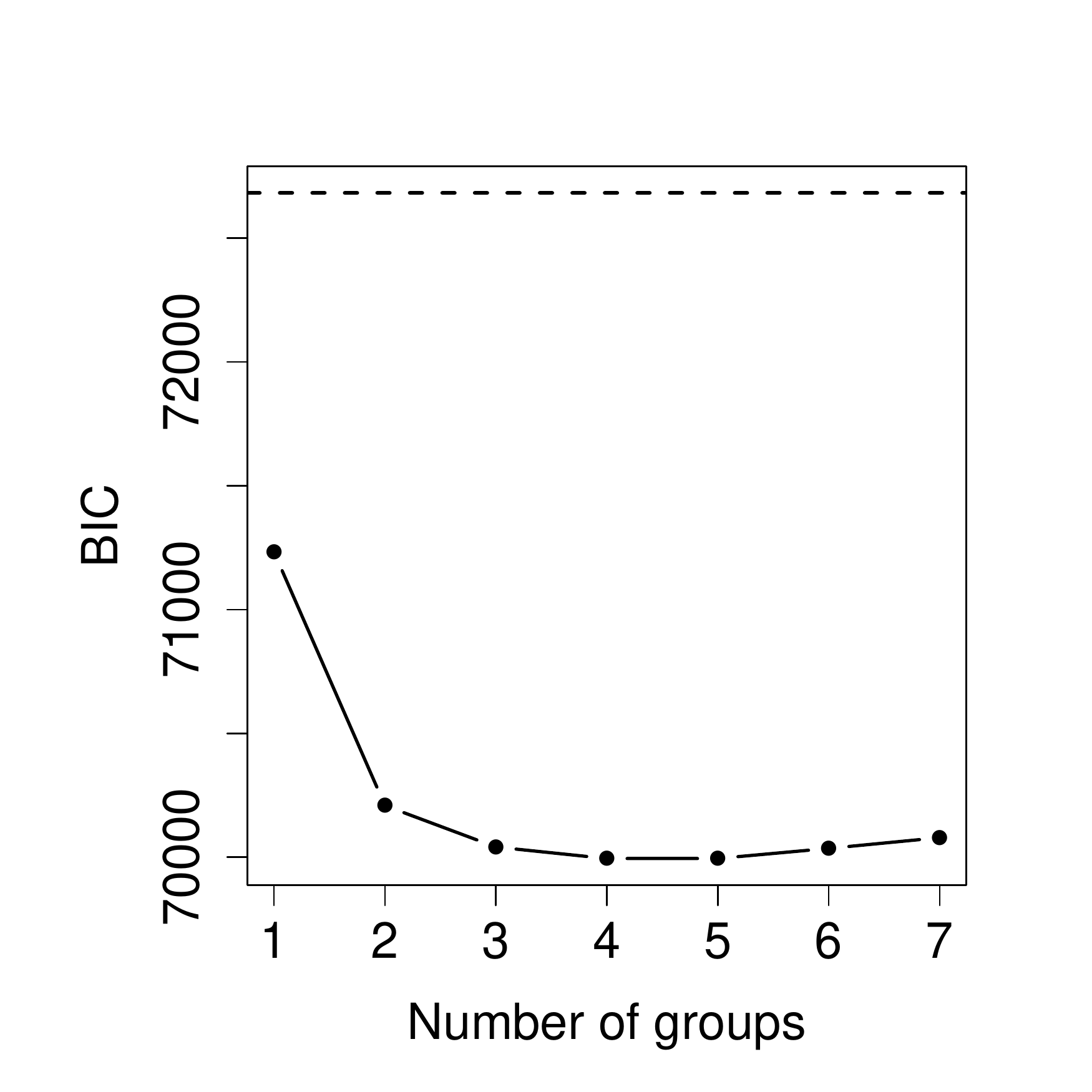}
        \caption{BIC values for the optimal panel fits as a function of different group sizes (solid line). The horizontal dashed line corresponds to the BIC of the local model, i.e.\ when $G=N=127$.}
        \label{mintemp_bic}
\end{figure}

\begin{figure}[htbp]
\begin{center}
       \includegraphics[width=4in]{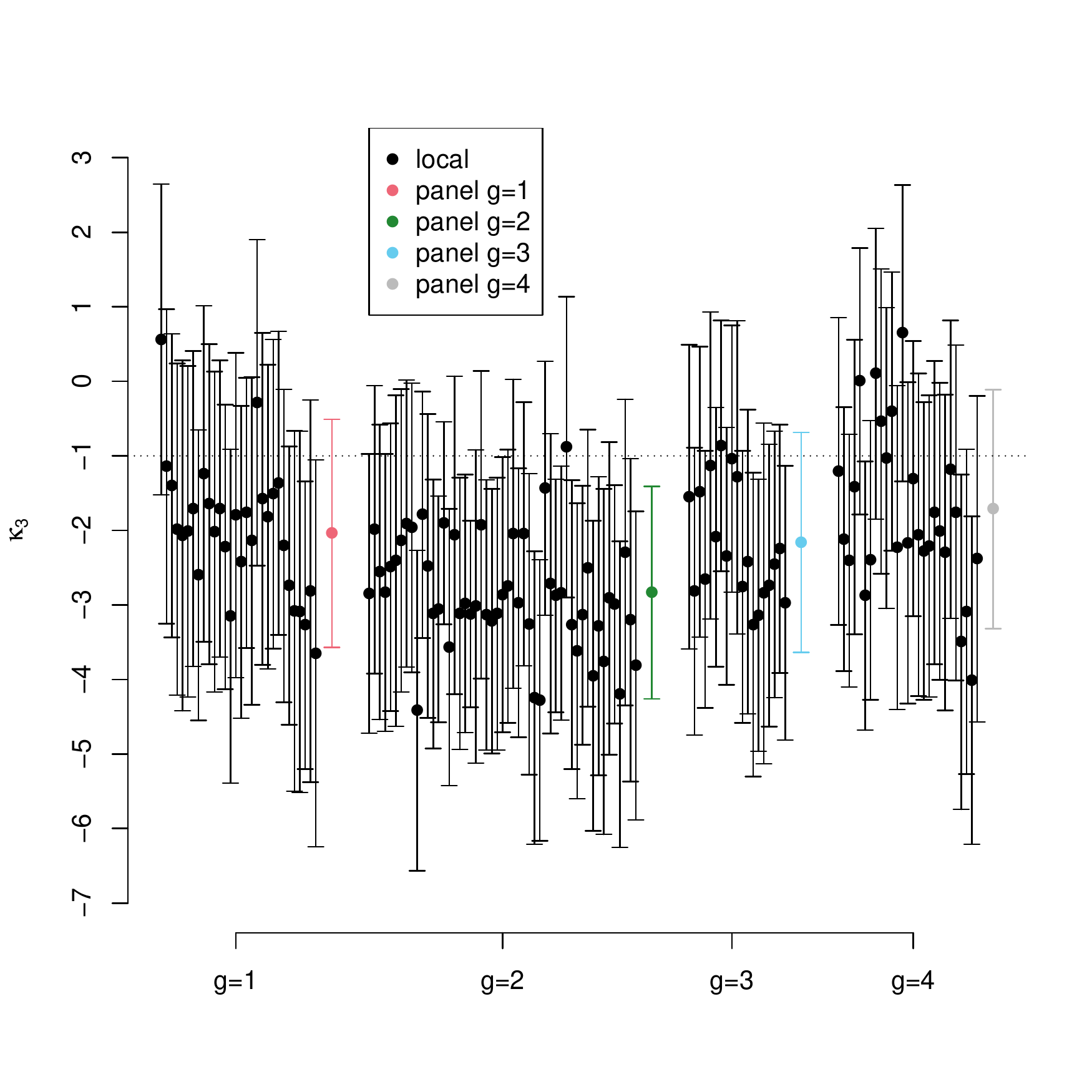}
        \null{\vspace{-0.4in}}
\end{center}
        \caption{Effect of global anomaly on negative annual daily
            minimum temperature based on GEV model fitted to data at each
            of the 127 locations (black, plotted by group)
            and panel estimates (color). 
            Grouping and colors shown
            in Figure \ref{Fig:MinTempBIC}.
            Horizontal dotted line indicates a 1:1 mean annual
            global anomaly
            to mean annual minimum temperature increase.}
        \label{Fig:MinTempGlobA}
\end{figure}

Figure \ref{mintemp_bic} shows BIC values of optimal panel fits 
for different group sizes.
Table~\ref{Tab:MinTempParms} shows the estimated parameters 
for panel model (\ref{eq:TempPanelGG}) with $G=1$ group
as well as the estimated
parameters 
with $G=G^*=4$, the optimal number of groups based on the BIC criterion.
Figure \ref{Fig:MinTempGlobA} shows the (local) estimates of
$\kappa_{3,(g)}$ when a GEV model with covariates as in (\ref{eq:TempPanelGG})
is fitted to data at each of the 127 
locations, i.e., $G=127$ groups,
as well as the estimates based on the GEV panel model 
with $G^*=4$.
The panel estimates with $G^*=4$ have better precision
by pooling information across all stations within a group,
showing 29, 39, 37 and 28\% mean
reduction in estimated standard errors compared to local estimates
for groups $g=1$ to $g=4$, respectively.
The panel fit allows us to infer
that a 1$\null^\circ$C increase in the annual global land anomaly results
in mean increases in the annual daily
minimum temperature between 1.7 and 2.8$\null^\circ$C in the area under
study. Figure~\ref{Fig:MinTempBIC} shows the panel groupings.  Largest
mean increases in annual daily
minimum temperature are predominantly in the western-most states (Nebraska, 
Kansas, Iowa and Missouri). 
The panel fit with $G=1$ in Table \ref{Tab:MinTempParms}
yields $\hat{\kappa}_{3,(g)} = -2.2\null^\circ$C for the entire area under study,
but the qq-plots at each station (not shown) for this $G=1$ model are quite poor and
inference is questionable.
Station-wise qq-plots (not shown) based on the $G^*=4$ panel model 
are very good, so the panel model
allows us to confidently infer that a $1^\circ$C increase in the
annual global anomaly results in mean increases 
to annual daily minimum temperature in
some U.S.\ Midwest regions that are up to almost three-fold that amount.
Continued exacerbated local effects of likely global increases
would be particularly problematic for these important 
crop-producing U.S.\ Midwest
states where temperature-driven crop yield variability is already
well documented
\citep{kukal2018, petersen2019}. 
Changes to the annual global land anomaly do not have a significant
effect on the variability of annual daily minimum temperatures 
as $\gamma_{3,(g)}$ estimates in
Table \ref{Tab:MinTempParms} are small when compared to their
associated standard errors.

\begin{table}[htbp]
        \centering
        \caption{Parameter estimates of the models for negative annual
                     daily minimum temperatures:
                     single group $G=1$ and optimal group structure
                     $\boldsymbol{\widehat{\tau}}^*$ from the EM
                     algorithm.} 
	\begin{tabular}{cccccc}
		\toprule
		\textit{Parameter} & $G=1$ & \multicolumn{4}{c}{$\boldsymbol{\widehat{\tau}}^*$}    \\
		\midrule
		&                    & $g=1$ & $g=2$ & $g=3$ & $g=4$ \\
		\cmidrule{3-6}       
$\kappa_{0,(g)}$     & -23.7  & -24.2 & -22.9 & -22.1 & -22.6 \\
                     & (0.5) & (0.8) & (0.5) & (0.5) & (0.6) \\
$\kappa_{1,(g)}$     &  3.7 &   3.3 &   2.9 &   4.0  &   5.1 \\
                     & (0.5) & (0.5) & (0.5) & (0.8) & (0.5) \\
$\kappa_{2,(g)}$     &  17.7 &  18.3  & 18.4 &  17.8 &  10.7 \\
                     & (0.4) & (0.8) & (0.5) & (0.5) & (0.8) \\
$\kappa_{3,(g)}$     &  -2.2 &  -2.0 &  -2.8  & -2.2  &  -1.7 \\
                     & (0.7) & (0.8) & (0.7) & (0.8) & (0.8) \\
$\gamma_{0,(g)}$     &  1.59  &  1.87 &   1.55  &  1.53  & 1.64 \\
                     & (0.06) & (0.12) & (0.08) & (0.07) & (0.08) \\
$\gamma_{1,(g)}$     &  -0.09 &  0.13  & -0.22  &  0.03  & -0.01 \\
                     & (0.06) & (0.08) & (0.09) & (0.15) & (0.08) \\
$\gamma_{2,(g)}$     &  -0.30 &  -0.9 &  -0.4 & -0.4  & -0.4 \\
		     & (0.09) & (0.2) & (0.1) & (0.1) & (0.1) \\
$\gamma_{3,(g)}$     &   0.04 &  0.10 &  0.10 & 0.03  & 0.02  \\
		     & (0.07) & (0.09) & (0.08) & (0.09) & (0.1) \\
$\delta_{0,(g)}$     &  -0.24  & -0.25  & -0.27 &  -0.23  & -0.25 \\
		     & (0.01) & (0.02) & (0.02) & (0.02) & (0.02) \\
		\bottomrule
        \end{tabular}
        \label{Tab:MinTempParms}
\end{table}

\section{Flood risk assessment}
\label{Sec:EA3}

Floods are major natural hazards that threaten human lives and cause huge damages to the environment and the economy. Effective flood protection is therefore crucial and this requires an accurate assessment of the risk related to high river flows. 

The GEV distribution $H$ in~\eqref{gev_form} is a widely used model for yearly maxima of river discharges thanks to its mathematical justification and good properties in applications \citep[e.g.][]{kat2002}. At a gauging station with many consecutive years of observations, the parameters $\mu$, $\sigma$ and $\xi$ of this distribution family can be fitted locally as in~\eqref{gev_mle}, that is, using only data from this particular station.

\begin{figure}[htbp]
  \centering
  \includegraphics[width=0.7\linewidth]{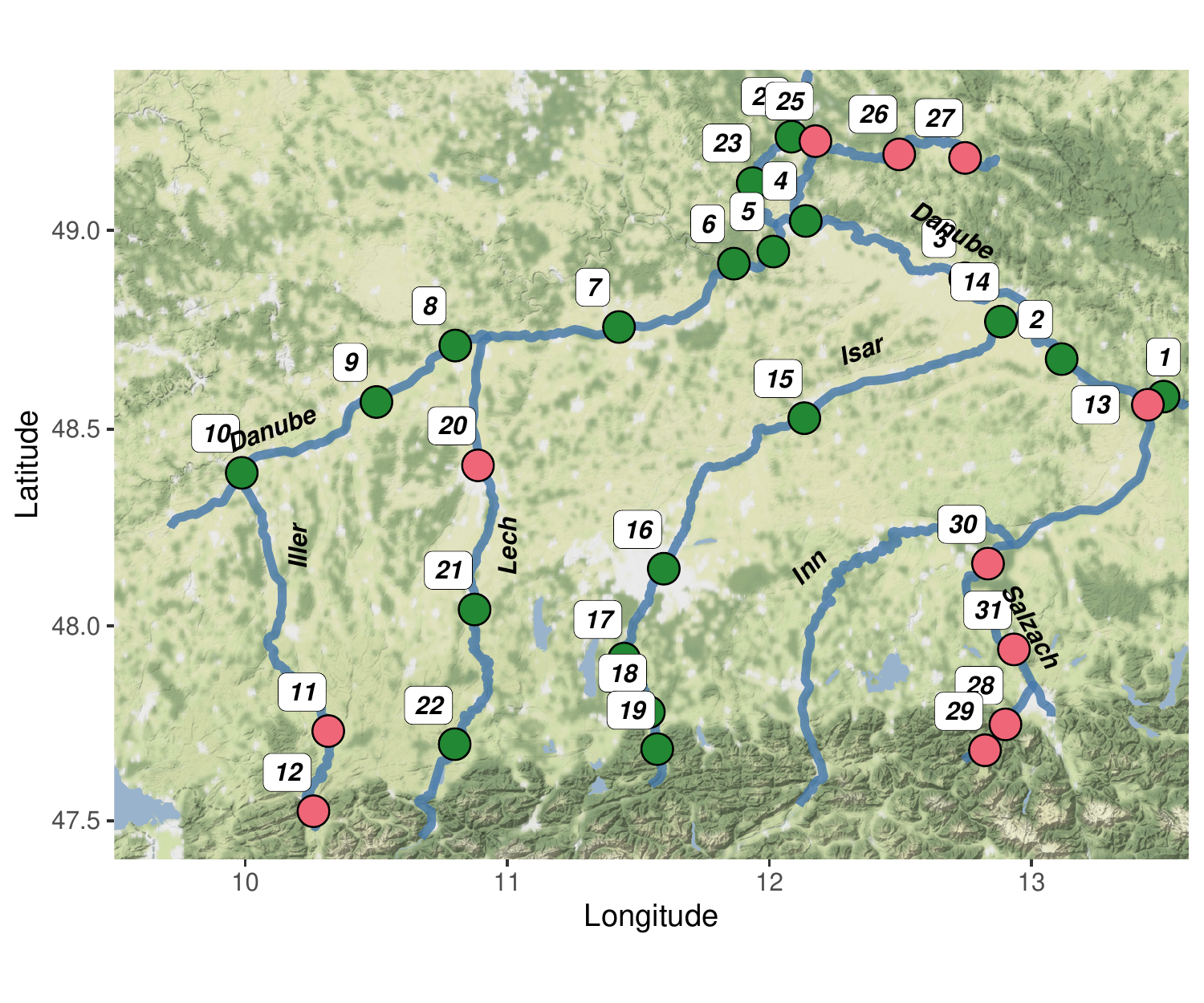}
        \caption{Upper Danube river basin and 31 gauging stations. The $G^*=2$ different colors of the stations correspond to the optimal grouping for the GEV panel.}
  	\label{danube_stations}
\end{figure}

In many hydrological applications, analysis of flood risk is required at stations with little or no data, and in such cases estimates of the return level $RL^S(\mu,\sigma,\xi)$ for long return periods $S$ based on the the local fit may exhibit huge variances. Regionalization is a common alternative way of estimating such high quantiles. It relies on identifying groups of stations that are similar to each other and on sharing their information on extreme flows to obtain more accurate estimates. There exists a vast literature that contains many different methods to construct such groupings and different ways of sharing the information \citep[e.g.,][]{bur1990, mer2005, asa2018}. For a fixed grouping, a widely used model for the $i$th station is
\begin{align}
  \notag\log(\mu_{i,t}(\tau_i))  &=  \boldsymbol{\kappa}_{(\tau_i)}^{\top} \log(\mathbf{X}_{i,t}),\\
  \label{cov_model}\log(\sigma_{i,t}(\tau_i))  &= \boldsymbol{\gamma}_{(\tau_i)}^{\top} \log(\mathbf{X}_{i,t}),\\
  \notag\xi_{i,t}(\tau_i)  &= \xi_{(\tau_i)},
\end{align}
with $\tau_i\in\left\{1,\dots,G\right\}$. We consider yearly maxima of river flow data $Y_{i,t}$, $i=1,\dots, N$, $t=1,\dots, T$, at $N=31$ stations in the upper Danube catchment during $T=50$ years; see Figure~\ref{danube_stations} for the river network and the locations of the gauging stations. This data set has been used in \cite{asadi2015extremes}, \cite{gne2021}, \cite{mha2020}, \cite{roe2021} and \cite{eng2018} for univariate and multivariate extreme value analyses. In addition to the river flow measurements, for each observation $Y_{i,t}$ we use a corresponding covariate vector $\mathbf{X}_{i,t}$ that contains the latitude of the station, and the size, the mean altitude and the mean slope of the corresponding sub-catchment. Since the data are yearly maxima, we can interpret this as a grouped panel GEV regression as in~\eqref{grouped_panel}, and we see that the covariate vectors are time-invariant, that is, $\mathbf{X}_{i,t} \equiv \mathbf{X}_{i}$.

For risk assessment at each of the $N=31$ stations there are several possibilities. One may locally fit at each station separately a GEV distribution. As discussed above, this becomes infeasible or highly sub-optimal if the data record is too short at some stations. To illustrate this, we choose six stations and randomly delete $80\%$ of their $T=50$ observations. 
A local fit at each station can be seen as a grouped panel with $G = N = 31$
groups, that is, every station is in its own group and no information is shared.
To borrow information across the $31$ stations, one can use a covariate model as in~\eqref{cov_model} for all stations simultaneously. This corresponds to a panel model with $G=1$ group only. Since the effect of the covariates might however not be the same for all stations, a regionalization approach with a good group assigment of the stations should provide a superior fit. \cite{asadi2015extremes} choose such a grouping with $G=4$ groups in an \textit{ad hoc} way and fit model~\eqref{cov_model} to the data.

\begin{figure}[htbp]
  \centering
		\includegraphics[width=0.49\linewidth]{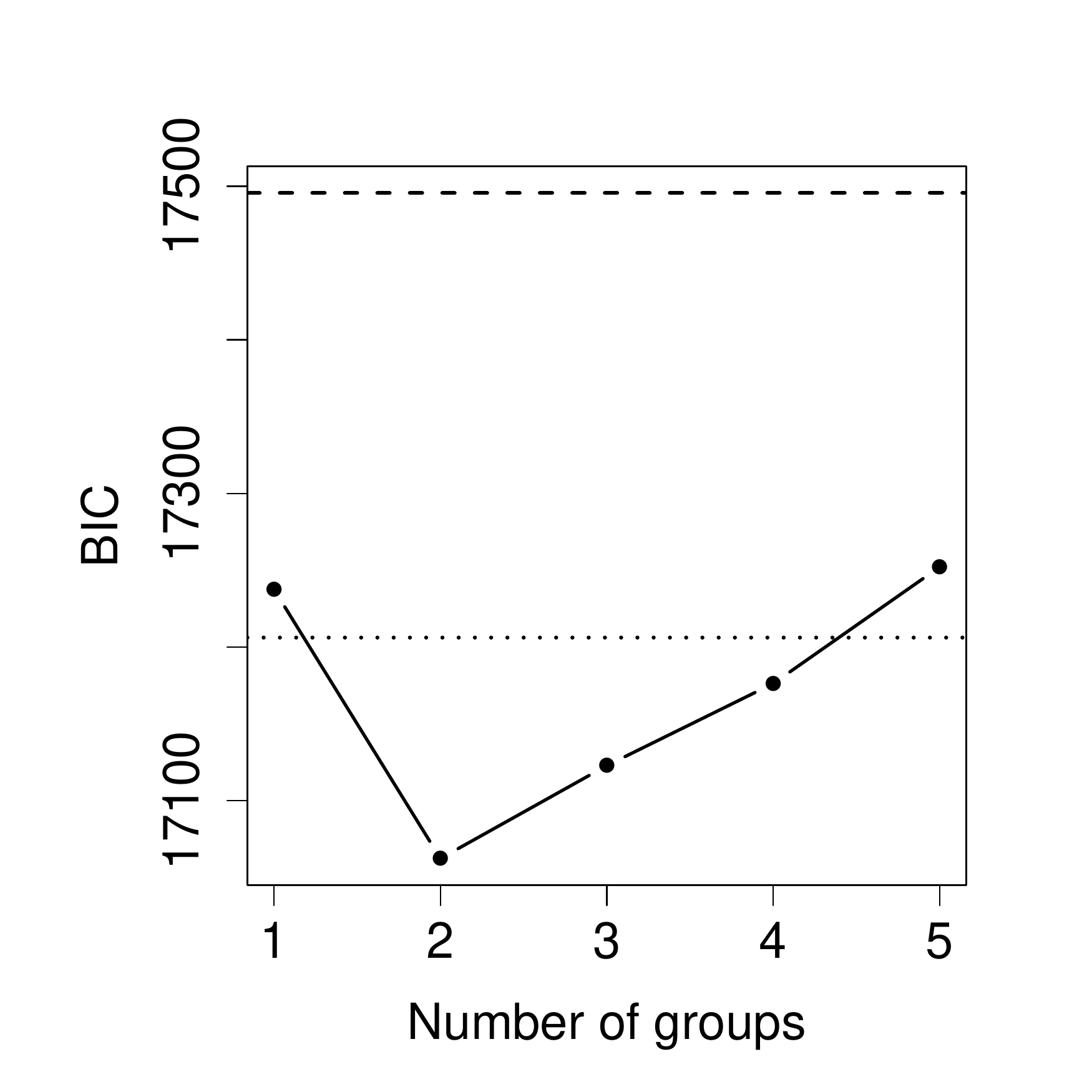}
        \caption{BIC values for the optimal panel fit as a function of different number of groups (solid line). The horizontal lines correspond to the BIC of the local model (dashed line) and the model in \cite{asadi2015extremes} (dotted line), respectively.}
	\label{danube_aicbic}
\end{figure}

Our grouped panel approach allows us to find the optimal group assignments  as described in Section~\ref{Sec:GroupPanel}. The number of groups for this data set is $G^*=2$ and is chosen as the panel that minimizes the BIC; see Figure~\ref{danube_aicbic} for the BIC for different numbers of groups. The optimal number of groups chosen by our algorithm is smaller than in \cite{asadi2015extremes} and therefore allows for more efficient pooling of information. Note that even if we consider a fixed group size $G=4$, the group assignments of our panel with four groups differ from the grouping in their approach.  While they use an \textit{ad hoc} grouping based on domain knowledge, we optimize our assignment in a purely data-driven way according to the second step in the EM algorithm in Section~\ref{Sec:GroupPanel}.

\begin{figure}[htbp]
  \hspace*{-3em}
  \includegraphics[width=1\linewidth]{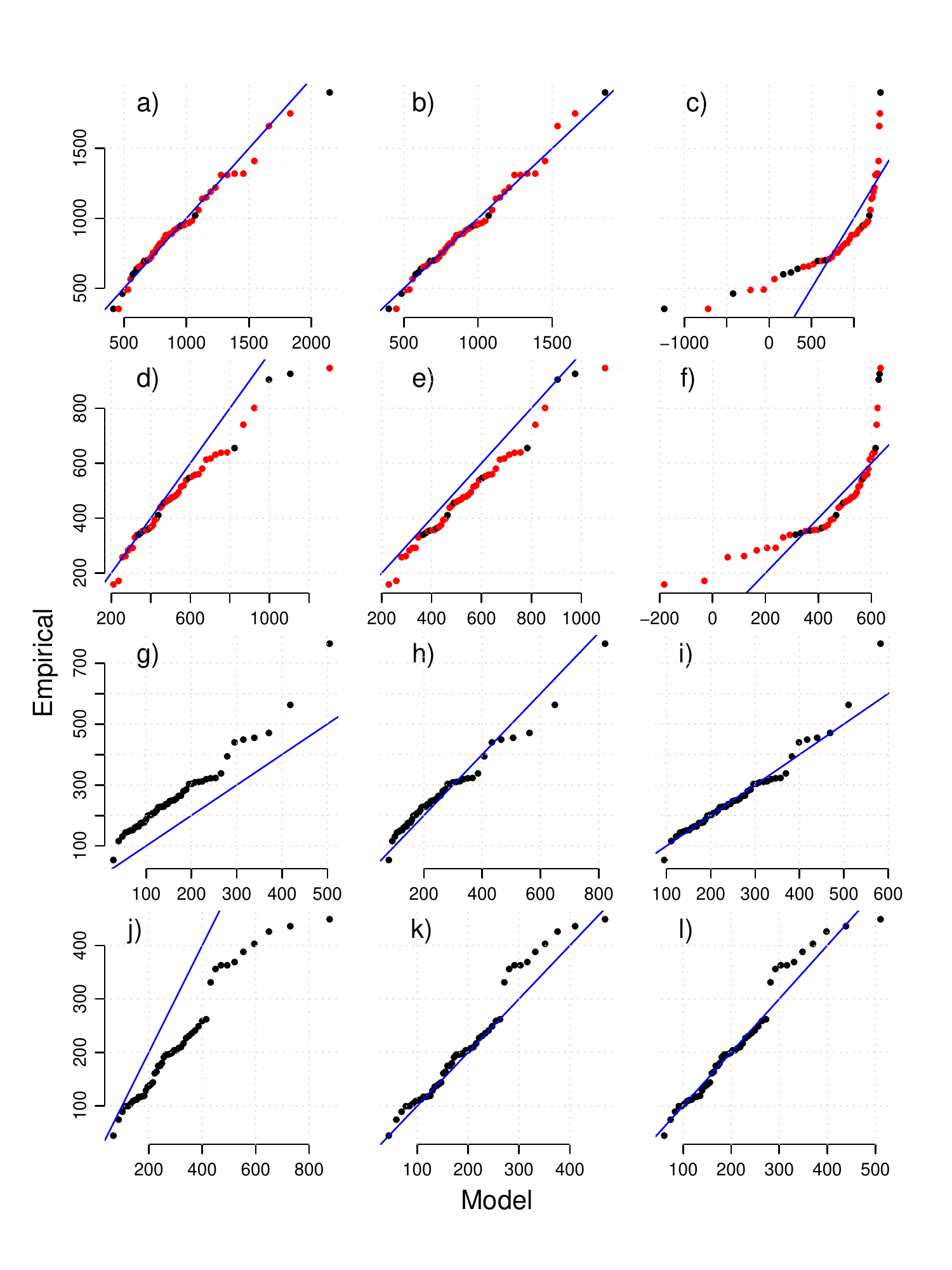} 
  \vspace{-0.4in}
  \caption{The rows correspond to QQ-plots of four stations: fit with $G=1$ group (left), fit with $G^*=2$ groups (center) and local fit (right). Red points are missing data that are not used for fitting in any of the models.}
    \label{danube_qq}
\end{figure}

Figure~\ref{danube_qq} shows the QQ-plots of the different fitted models for four exemplary stations, where two of them have missing data. It can be seen that the local fit (figures on the right) performs well for stations where enough data are available, but naturally it is not able to capture well the tail if data at a station are scarce. For the other boundary case of a global fit with only one group, $G=1$, the left-hand side of Figure~\ref{danube_qq} shows that such a model is not flexible enough to model the extremes at all locations well. Our optimized panel with data-driven group assignments is shown in the center column of Figure~\ref{danube_qq}.  We see that it combines the advantages of the local and the global fits. It is flexible enough to model the tail at all stations sufficiently well. Moreover, the fact of pooling the information from all stations in a group helps to obtain a good fit even for stations with a lot of missing data.
The BIC values in Figure~\ref{danube_aicbic} provide further evidence that our optimized grouped panel GEV regression performs better than the local and global models, and the model of \cite{asadi2015extremes}.

Figure~\ref{danube_stations} shows the final group assignments of our optimal panel model in two different colors. We recognize a clear spatial pattern in the sense that stations on the same river tend to fall into the same group. This is astonishing since our method does not enforce this in any way. This shows the big advantage of our methodology, namely, that it does not require domain knowledge to produce sensible panel groupings.

\section{Discussion}
\label{Sec:End}
Sections \ref{Sec:EA1} to \ref{Sec:EA3} present three applications demonstrating the practical usefulness of our model in very diverse settings, but the benefits of properly grouping individuals to estimate extreme events extend to other situations. We mention a few of the many applications: applying our grouped panel GEV regression model to i) annual maximum sub-hourly precipitation would reduce the estimation error of design storms at more recent gauging sites where only short series of precipitation data are available \citep{alila1999, buishand1991}; ii) annual maximum three-day snow fall depth data would reduce the estimation error of the extreme return levels required in avalanche hazard mapping \citep{bocchiola2008}; iii) annual fastest-mile wind speed data would reduce the estimation error of at-site minimum design wind loads \citep{cheng1998}.

\red{We present our approach and applications based on the GEV regression model where extremes are defined as the maximum (or minimum) observation from blocks of equal size. A different modeling framework for extremes is the peaks-over-threshold method that uses the generalized Pareto (GP) distribution \citep{pic1975}. In the Appendix we outline how such a grouped panel GP regression model can be implemented.}

The parameters in our grouped panel GEV or GP model are linear functions of the covariates. If more flexibility is needed, it is straightforward to generalize our approach to other forms of regression functions, such as additive models \citep{chavez2005generalized}, neural networks \citep{can2010, vel2021} or random forests \citep{gne2022}. \red{Variable selection through lasso-type techniques is another possible extension \citep{de2021extreme}.} 

\red{Future research may also extend our methodological framework to modeling of panels of multivariate extremes where the dependence structure of the maxima depends on a vector of covariates as in \cite{de2014spectral}. Latent groups with different dependence regimes could then be defined and our EM algorithm could be extended to estimate the regression parameters and group allocations in the dependence structure following the lines of \cite{oh2020dynamic}.}

\bibliographystyle{imsart-nameyear}
\bibliography{panel}

\begin{thebibliography}{70}

\bibitem[\protect\citeauthoryear{Alila}{1999}]{alila1999}
\begin{barticle}[author]
\bauthor{\bsnm{Alila},~\bfnm{Younes}\binits{Y.}}
(\byear{1999}).
\btitle{A hierarchical approach for the regionalization of precipitation annual
  maxima in {C}anada}.
\bjournal{Journal of Geophysical Research}
\bvolume{104}
\bpages{31,645--31,655}.
\end{barticle}
\endbibitem

\bibitem[\protect\citeauthoryear{Asadi, Davison and
  Engelke}{2015}]{asadi2015extremes}
\begin{barticle}[author]
\bauthor{\bsnm{Asadi},~\bfnm{Peiman}\binits{P.}},
  \bauthor{\bsnm{Davison},~\bfnm{Anthony~C.}\binits{A.~C.}} \AND
  \bauthor{\bsnm{Engelke},~\bfnm{Sebastian}\binits{S.}}
(\byear{2015}).
\btitle{Extremes on river networks}.
\bjournal{The Annals of Applied Statistics}
\bvolume{9}
\bpages{2023--2050}.
\end{barticle}
\endbibitem

\bibitem[\protect\citeauthoryear{Asadi, Engelke and Davison}{2018}]{asa2018}
\begin{barticle}[author]
\bauthor{\bsnm{Asadi},~\bfnm{Peiman}\binits{P.}},
  \bauthor{\bsnm{Engelke},~\bfnm{Sebastian}\binits{S.}} \AND
  \bauthor{\bsnm{Davison},~\bfnm{Anthony~C.}\binits{A.~C.}}
(\byear{2018}).
\btitle{Optimal regionalization of extreme value distributions for flood
  estimation}.
\bjournal{Journal of Hydrology}
\bvolume{556}
\bpages{182--193}.
\end{barticle}
\endbibitem

\bibitem[\protect\citeauthoryear{Bali}{2003}]{bali2003extreme}
\begin{barticle}[author]
\bauthor{\bsnm{Bali},~\bfnm{Turan~G}\binits{T.~G.}}
(\byear{2003}).
\btitle{An extreme value approach to estimating volatility and value at risk}.
\bjournal{The Journal of Business}
\bvolume{76}
\bpages{83--108}.
\end{barticle}
\endbibitem

\bibitem[\protect\citeauthoryear{Barndorff-Nielsen, Kinnebrouk and
  Shephard}{2010}]{barn2010}
\begin{binbook}[author]
\bauthor{\bsnm{Barndorff-Nielsen},~\bfnm{Ole~E.}\binits{O.~E.}},
  \bauthor{\bsnm{Kinnebrouk},~\bfnm{Silvia}\binits{S.}} \AND
  \bauthor{\bsnm{Shephard},~\bfnm{Neil}\binits{N.}}
(\byear{2010}).
\btitle{Measuring downside risk: realised semivariance},
In \bbooktitle{Volatility and Time Series Econometrics: Essays in Honor of
  Robert F. Engle}
\bedition{edited by {T}. {B}ollerslev, {J}. {R}ussell and {M}. {W}atson)} ed.
\bpages{117-136}.
\bpublisher{Oxford University Press}.
\end{binbook}
\endbibitem

\bibitem[\protect\citeauthoryear{Bee, Dupuis and
  Trapin}{2019}]{bee2019realized}
\begin{barticle}[author]
\bauthor{\bsnm{Bee},~\bfnm{Marco}\binits{M.}},
  \bauthor{\bsnm{Dupuis},~\bfnm{Debbie~J}\binits{D.~J.}} \AND
  \bauthor{\bsnm{Trapin},~\bfnm{Luca}\binits{L.}}
(\byear{2019}).
\btitle{Realized Peaks over Threshold: a Time-Varying Extreme Value Approach
  with High-Frequency-Based Measures}.
\bjournal{Journal of Financial Econometrics}
\bvolume{17}
\bpages{254--283}.
\end{barticle}
\endbibitem

\bibitem[\protect\citeauthoryear{Bocchiola et~al.}{2008}]{bocchiola2008}
\begin{barticle}[author]
\bauthor{\bsnm{Bocchiola},~\bfnm{D}\binits{D.}},
  \bauthor{\bsnm{Bianchi~Janetti},~\bfnm{E}\binits{E.}},
  \bauthor{\bsnm{Gorni},~\bfnm{E}\binits{E.}},
  \bauthor{\bsnm{Marty},~\bfnm{C}\binits{C.}} \AND
  \bauthor{\bsnm{Sovilla},~\bfnm{B}\binits{B.}}
(\byear{2008}).
\btitle{Regional evaluation of three-day snow depth for avalanche hazard
  mapping in {S}witzerland}.
\bjournal{Natural Hazards and Earth System Sciences}
\bvolume{8}
\bpages{685--705}.
\end{barticle}
\endbibitem

\bibitem[\protect\citeauthoryear{B{\"u}cher and Segers}{2017}]{bue2017}
\begin{barticle}[author]
\bauthor{\bsnm{B{\"u}cher},~\bfnm{Axel}\binits{A.}} \AND
  \bauthor{\bsnm{Segers},~\bfnm{Johan}\binits{J.}}
(\byear{2017}).
\btitle{On the maximum likelihood estimator for the generalized extreme-value
  distribution}.
\bjournal{Extremes}
\bvolume{20}
\bpages{839--872}.
\end{barticle}
\endbibitem

\bibitem[\protect\citeauthoryear{B\"{u}cher and
  Segers}{2018}]{bucher2018maximum}
\begin{barticle}[author]
\bauthor{\bsnm{B\"{u}cher},~\bfnm{Axel}\binits{A.}} \AND
  \bauthor{\bsnm{Segers},~\bfnm{Johan}\binits{J.}}
(\byear{2018}).
\btitle{Maximum likelihood estimation for the {F}r\'{e}chet distribution based
  on block maxima extracted from a time series}.
\bjournal{Bernoulli}
\bvolume{24}
\bpages{1427--1462}.
\bmrnumber{3706798}
\end{barticle}
\endbibitem

\bibitem[\protect\citeauthoryear{Buishand}{1991}]{buishand1991}
\begin{barticle}[author]
\bauthor{\bsnm{Buishand},~\bfnm{T.~A.}\binits{T.~A.}}
(\byear{1991}).
\btitle{Extreme rainfall estimation by combining data from several sites}.
\bjournal{Hydrological Sciences Journal}
\bvolume{36}
\bpages{345--365}.
\end{barticle}
\endbibitem

\bibitem[\protect\citeauthoryear{Burn}{1990}]{bur1990}
\begin{barticle}[author]
\bauthor{\bsnm{Burn},~\bfnm{D.~H.}\binits{D.~H.}}
(\byear{1990}).
\btitle{An appraisal of the “region of influence” approach to flood
  frequency analysis}.
\bjournal{Hydrological Sciences Journal}
\bvolume{35}
\bpages{149-165}.
\end{barticle}
\endbibitem

\bibitem[\protect\citeauthoryear{Cameron and Trivedi}{2015}]{cameron2015count}
\begin{bincollection}[author]
\bauthor{\bsnm{Cameron},~\bfnm{Colin~A.}\binits{C.~A.}} \AND
  \bauthor{\bsnm{Trivedi},~\bfnm{Pravin~K.}\binits{P.~K.}}
(\byear{2015}).
\btitle{Count panel data}.
In \bbooktitle{The Oxford Handbook of Panel Data}
(\beditor{\bfnm{Badi~H.}\binits{B.~H.}~\bsnm{Baltagi}}, ed.)
\bchapter{8}
\bpublisher{Oxford University Press}, \baddress{Oxford}.
\end{bincollection}
\endbibitem

\bibitem[\protect\citeauthoryear{Cannon}{2010}]{can2010}
\begin{barticle}[author]
\bauthor{\bsnm{Cannon},~\bfnm{Alex~J.}\binits{A.~J.}}
(\byear{2010}).
\btitle{A flexible nonlinear modelling framework for nonstationary generalized
  extreme value analysis in hydroclimatology}.
\bjournal{Hydrological Processes}
\bvolume{24}
\bpages{673-685}.
\end{barticle}
\endbibitem

\bibitem[\protect\citeauthoryear{Carreau, Naveau and
  Neppel}{2017}]{Carreau2017}
\begin{barticle}[author]
\bauthor{\bsnm{Carreau},~\bfnm{J.}\binits{J.}},
  \bauthor{\bsnm{Naveau},~\bfnm{P.}\binits{P.}} \AND
  \bauthor{\bsnm{Neppel},~\bfnm{L.}\binits{L.}}
(\byear{2017}).
\btitle{Partitioning into hazard subregions for regional peaks-over-threshold
  modeling of heavy precipitation}.
\bjournal{Water Resources Research}
\bvolume{53}
\bpages{4407--4426}.
\bdoi{10.1002/2017WR020758}
\end{barticle}
\endbibitem

\bibitem[\protect\citeauthoryear{Chandler and
  Bate}{2007}]{chandler2007inference}
\begin{barticle}[author]
\bauthor{\bsnm{Chandler},~\bfnm{Richard~E}\binits{R.~E.}} \AND
  \bauthor{\bsnm{Bate},~\bfnm{Steven}\binits{S.}}
(\byear{2007}).
\btitle{Inference for clustered data using the independence loglikelihood}.
\bjournal{Biometrika}
\bvolume{94}
\bpages{167--183}.
\end{barticle}
\endbibitem

\bibitem[\protect\citeauthoryear{Chavez-Demoulin and
  Davison}{2005}]{chavez2005generalized}
\begin{barticle}[author]
\bauthor{\bsnm{Chavez-Demoulin},~\bfnm{Val{\'e}rie}\binits{V.}} \AND
  \bauthor{\bsnm{Davison},~\bfnm{Anthony~C}\binits{A.~C.}}
(\byear{2005}).
\btitle{Generalized additive modelling of sample extremes}.
\bjournal{Journal of the Royal Statistical Society: Series C}
\bvolume{54}
\bpages{207--222}.
\end{barticle}
\endbibitem

\bibitem[\protect\citeauthoryear{Cheng}{1998}]{cheng1998}
\begin{barticle}[author]
\bauthor{\bsnm{Cheng},~\bfnm{Edmond D.~H.}\binits{E.~D.~H.}}
(\byear{1998}).
\btitle{Macroscopic extreme wind regionalization}.
\bjournal{Journal of Wind Engineering and Industrial Aerodynamics}
\bvolume{77-78}
\bpages{13--21}.
\end{barticle}
\endbibitem

\bibitem[\protect\citeauthoryear{Coles}{2001}]{coles2001introduction}
\begin{bbook}[author]
\bauthor{\bsnm{Coles},~\bfnm{Stuart}\binits{S.}}
(\byear{2001}).
\btitle{An Introduction to Statistical Modeling of Extreme Values}.
\bpublisher{Springer}.
\end{bbook}
\endbibitem

\bibitem[\protect\citeauthoryear{Davison, Padoan and
  Ribatet}{2012}]{davison2012statistical}
\begin{barticle}[author]
\bauthor{\bsnm{Davison},~\bfnm{Anthony~C}\binits{A.~C.}},
  \bauthor{\bsnm{Padoan},~\bfnm{Simone~A}\binits{S.~A.}} \AND
  \bauthor{\bsnm{Ribatet},~\bfnm{Mathieu}\binits{M.}}
(\byear{2012}).
\btitle{Statistical modeling of spatial extremes}.
\bjournal{Statistical Science}
\bvolume{27}
\bpages{161--186}.
\end{barticle}
\endbibitem

\bibitem[\protect\citeauthoryear{de~Carvalho and
  Davison}{2014}]{de2014spectral}
\begin{barticle}[author]
\bauthor{\bparticle{de} \bsnm{Carvalho},~\bfnm{Miguel}\binits{M.}} \AND
  \bauthor{\bsnm{Davison},~\bfnm{Anthony~C}\binits{A.~C.}}
(\byear{2014}).
\btitle{Spectral density ratio models for multivariate extremes}.
\bjournal{Journal of the American Statistical Association}
\bvolume{109}
\bpages{764--776}.
\end{barticle}
\endbibitem

\bibitem[\protect\citeauthoryear{de~Carvalho et~al.}{2021}]{de2021extreme}
\begin{barticle}[author]
\bauthor{\bparticle{de} \bsnm{Carvalho},~\bfnm{Miguel}\binits{M.}},
  \bauthor{\bsnm{Pereira},~\bfnm{Soraia}\binits{S.}},
  \bauthor{\bsnm{Pereira},~\bfnm{Paula}\binits{P.}} \AND
  \bauthor{\bparticle{de} \bsnm{Zea~Bermudez},~\bfnm{Patr{\'\i}cia}\binits{P.}}
(\byear{2021}).
\btitle{An Extreme Value Bayesian Lasso for the Conditional Left and Right
  Tails}.
\bjournal{Journal of Agricultural, Biological and Environmental Statistics}
\bpages{1--18}.
\end{barticle}
\endbibitem

\bibitem[\protect\citeauthoryear{Dombry}{2015}]{dom2015}
\begin{barticle}[author]
\bauthor{\bsnm{Dombry},~\bfnm{Cl\'{e}ment}\binits{C.}}
(\byear{2015}).
\btitle{Existence and consistency of the maximum likelihood estimators for the
  extreme value index within the block maxima framework}.
\bjournal{Bernoulli}
\bvolume{21}
\bpages{420--436}.
\bmrnumber{3322325}
\end{barticle}
\endbibitem

\bibitem[\protect\citeauthoryear{Dombry and Ferreira}{2019}]{dom2019}
\begin{barticle}[author]
\bauthor{\bsnm{Dombry},~\bfnm{Cl\'{e}ment}\binits{C.}} \AND
  \bauthor{\bsnm{Ferreira},~\bfnm{Ana}\binits{A.}}
(\byear{2019}).
\btitle{Maximum likelihood estimators based on the block maxima method}.
\bjournal{Bernoulli}
\bvolume{25}
\bpages{1690--1723}.
\bmrnumber{3961227}
\end{barticle}
\endbibitem

\bibitem[\protect\citeauthoryear{Embrechts, Kl{\"u}ppelberg and
  Mikosch}{1997}]{embrechts1997modelling}
\begin{bbook}[author]
\bauthor{\bsnm{Embrechts},~\bfnm{Paul}\binits{P.}},
  \bauthor{\bsnm{Kl{\"u}ppelberg},~\bfnm{Claudia}\binits{C.}} \AND
  \bauthor{\bsnm{Mikosch},~\bfnm{Thomas}\binits{T.}}
(\byear{1997}).
\btitle{Modelling Extremal Events: for Insurance and Finance}.
\bpublisher{Springer Science \& Business Media}.
\end{bbook}
\endbibitem

\bibitem[\protect\citeauthoryear{Engelke and Hitz}{2020}]{eng2018}
\begin{barticle}[author]
\bauthor{\bsnm{Engelke},~\bfnm{S.}\binits{S.}} \AND
  \bauthor{\bsnm{Hitz},~\bfnm{A.~S.}\binits{A.~S.}}
(\byear{2020}).
\btitle{Graphical models for extremes (with discussion)}.
\bjournal{J. R. Stat. Soc. Ser. B Stat. Methodol.}
\bvolume{82}
\bpages{871--932}.
\end{barticle}
\endbibitem

\bibitem[\protect\citeauthoryear{Garc\'ia et~al.}{2015}]{garcia2015}
\begin{barticle}[author]
\bauthor{\bsnm{Garc\'ia},~\bfnm{Guillermo}\binits{G.}},
  \bauthor{\bsnm{Dreccer},~\bfnm{M.}\binits{M.}},
  \bauthor{\bsnm{Miralles},~\bfnm{Daniel}\binits{D.}} \AND
  \bauthor{\bsnm{Serrago},~\bfnm{Rom\'an}\binits{R.}}
(\byear{2015}).
\btitle{High night temperatures during grain number determination reduce wheat
  and barley grain yield: a field study}.
\bjournal{Global Change Biology}
\bvolume{21}
\bpages{4153--4164}.
\bdoi{10.1111/gcb.13009}
\end{barticle}
\endbibitem

\bibitem[\protect\citeauthoryear{Gnecco, Terefe and Engelke}{2022}]{gne2022}
\begin{bmisc}[author]
\bauthor{\bsnm{Gnecco},~\bfnm{Nicola}\binits{N.}},
  \bauthor{\bsnm{Terefe},~\bfnm{Edossa~Merga}\binits{E.~M.}} \AND
  \bauthor{\bsnm{Engelke},~\bfnm{Sebastian}\binits{S.}}
(\byear{2022}).
\btitle{Extremal Random Forests}.
\bdoi{10.48550/ARXIV.2201.12865}
\end{bmisc}
\endbibitem

\bibitem[\protect\citeauthoryear{Gnecco et~al.}{2021}]{gne2021}
\begin{barticle}[author]
\bauthor{\bsnm{Gnecco},~\bfnm{Nicola}\binits{N.}},
  \bauthor{\bsnm{Meinshausen},~\bfnm{Nicolai}\binits{N.}},
  \bauthor{\bsnm{Peters},~\bfnm{Jonas}\binits{J.}} \AND
  \bauthor{\bsnm{Engelke},~\bfnm{Sebastian}\binits{S.}}
(\byear{2021}).
\btitle{{Causal discovery in heavy-tailed models}}.
\bjournal{The Annals of Statistics}
\bvolume{49}
\bpages{1755--1778}.
\end{barticle}
\endbibitem

\bibitem[\protect\citeauthoryear{Greene}{2009}]{greene2009discrete}
\begin{bincollection}[author]
\bauthor{\bsnm{Greene},~\bfnm{William}\binits{W.}}
(\byear{2009}).
\btitle{Discrete choice modeling}.
In \bbooktitle{Palgrave Handbook of Econometrics}
\bpages{473--556}.
\bpublisher{Springer}.
\end{bincollection}
\endbibitem

\bibitem[\protect\citeauthoryear{Gu and Volgushev}{2019}]{gu2019panel}
\begin{barticle}[author]
\bauthor{\bsnm{Gu},~\bfnm{Jiaying}\binits{J.}} \AND
  \bauthor{\bsnm{Volgushev},~\bfnm{Stanislav}\binits{S.}}
(\byear{2019}).
\btitle{Panel data quantile regression with grouped fixed effects}.
\bjournal{Journal of Econometrics}
\bvolume{213}
\bpages{68--91}.
\end{barticle}
\endbibitem

\bibitem[\protect\citeauthoryear{Hambuckers and
  Kneib}{2021}]{hambuckers2021smooth}
\begin{barticle}[author]
\bauthor{\bsnm{Hambuckers},~\bfnm{Julien}\binits{J.}} \AND
  \bauthor{\bsnm{Kneib},~\bfnm{Thomas}\binits{T.}}
(\byear{2021}).
\btitle{Smooth Transition Regression Models for Non-Stationary Extremes}.
\bjournal{Journal of Financial Econometrics}
\bvolume{nbab005}.
\end{barticle}
\endbibitem

\bibitem[\protect\citeauthoryear{Hansen et~al.}{2010}]{hansen2010}
\begin{barticle}[author]
\bauthor{\bsnm{Hansen},~\bfnm{J.}\binits{J.}},
  \bauthor{\bsnm{Ruedy},~\bfnm{R.}\binits{R.}},
  \bauthor{\bsnm{Sato},~\bfnm{M.}\binits{M.}} \AND
  \bauthor{\bsnm{Lo},~\bfnm{K.}\binits{K.}}
(\byear{2010}).
\btitle{Global surface temperature change}.
\bjournal{Reviews of Geophysics}
\bvolume{48}.
\bnote{RG4004}.
\end{barticle}
\endbibitem

\bibitem[\protect\citeauthoryear{Hsiao}{2007}]{hsiao2007panel}
\begin{barticle}[author]
\bauthor{\bsnm{Hsiao},~\bfnm{Cheng}\binits{C.}}
(\byear{2007}).
\btitle{Panel data analysis - advantages and challenges}.
\bjournal{Test}
\bvolume{16}
\bpages{1--22}.
\end{barticle}
\endbibitem

\bibitem[\protect\citeauthoryear{Hsiao}{2014}]{hsiao2014analysis}
\begin{bbook}[author]
\bauthor{\bsnm{Hsiao},~\bfnm{Cheng}\binits{C.}}
(\byear{2014}).
\btitle{Analysis of Panel Data}
\bvolume{54}.
\bpublisher{Cambridge University Press}.
\end{bbook}
\endbibitem

\bibitem[\protect\citeauthoryear{IPCC}{2008}]{ipcc}
\begin{bmisc}[author]
\bauthor{\bsnm{IPCC}}
(\byear{2008}).
\btitle{{C}limate change 2007. {S}ynthesis report. {C}ontribution of {W}orking
  {G}roups {I}, {II} and {III} to the {F}ourth {A}ssessment {R}eport of the
  {I}ntergovernmental {P}anel on {C}limate {C}hange {C}ore {W}riting {T}eam,
  eds. {R. K. P}achauri and {A. R}eisinger}.
\end{bmisc}
\endbibitem

\bibitem[\protect\citeauthoryear{Jurado, Ludvigson and
  Ng}{2015}]{jurado2015measuring}
\begin{barticle}[author]
\bauthor{\bsnm{Jurado},~\bfnm{Kyle}\binits{K.}},
  \bauthor{\bsnm{Ludvigson},~\bfnm{Sydney~C}\binits{S.~C.}} \AND
  \bauthor{\bsnm{Ng},~\bfnm{Serena}\binits{S.}}
(\byear{2015}).
\btitle{Measuring uncertainty}.
\bjournal{American Economic Review}
\bvolume{105}
\bpages{1177--1216}.
\end{barticle}
\endbibitem

\bibitem[\protect\citeauthoryear{Katz}{2013}]{katz2013statistical}
\begin{bincollection}[author]
\bauthor{\bsnm{Katz},~\bfnm{Richard~W}\binits{R.~W.}}
(\byear{2013}).
\btitle{Statistical methods for nonstationary extremes}.
In \bbooktitle{Extremes in a Changing Climate}
\bpages{15--37}.
\bpublisher{Springer}.
\end{bincollection}
\endbibitem

\bibitem[\protect\citeauthoryear{Katz, Parlange and Naveau}{2002}]{kat2002}
\begin{barticle}[author]
\bauthor{\bsnm{Katz},~\bfnm{R.~W.}\binits{R.~W.}},
  \bauthor{\bsnm{Parlange},~\bfnm{M.~B.}\binits{M.~B.}} \AND
  \bauthor{\bsnm{Naveau},~\bfnm{P.}\binits{P.}}
(\byear{2002}).
\btitle{Statistics of extremes in hydrology}.
\bjournal{Advances in Water Resources}
\bvolume{25}
\bpages{1287--1304}.
\end{barticle}
\endbibitem

\bibitem[\protect\citeauthoryear{Kukal and Irmak}{2018}]{kukal2018}
\begin{barticle}[author]
\bauthor{\bsnm{Kukal},~\bfnm{M.~S.}\binits{M.~S.}} \AND
  \bauthor{\bsnm{Irmak},~\bfnm{S.}\binits{S.}}
(\byear{2018}).
\btitle{Climate-Driven Crop Yield and Yield Variability and Climate Change
  Impacts on the {U.S.} {G}reat {P}lains Agricultural Production}.
\bjournal{Scientific Reports}
\bvolume{8}.
\bdoi{https://doi.org/10.1038/s41598-018-21848-2}
\end{barticle}
\endbibitem

\bibitem[\protect\citeauthoryear{Leadbetter, Lindgren and
  Rootz{\'e}n}{1983}]{leadbetter1983extremes}
\begin{bbook}[author]
\bauthor{\bsnm{Leadbetter},~\bfnm{Malcolm~R}\binits{M.~R.}},
  \bauthor{\bsnm{Lindgren},~\bfnm{Georg}\binits{G.}} \AND
  \bauthor{\bsnm{Rootz{\'e}n},~\bfnm{Holger}\binits{H.}}
(\byear{1983}).
\btitle{Extremes and Related Properties of Random Sequences and Processes}.
\bpublisher{Springer}.
\end{bbook}
\endbibitem

\bibitem[\protect\citeauthoryear{Massacci}{2017}]{massacci2017tail}
\begin{barticle}[author]
\bauthor{\bsnm{Massacci},~\bfnm{Daniele}\binits{D.}}
(\byear{2017}).
\btitle{Tail risk dynamics in stock returns: links to the macroeconomy and
  global markets connectedness}.
\bjournal{Management Science}
\bvolume{63}
\bpages{3072--3089}.
\end{barticle}
\endbibitem

\bibitem[\protect\citeauthoryear{McLachlan and Krishnan}{2008}]{McLachlanEM}
\begin{bbook}[author]
\bauthor{\bsnm{McLachlan},~\bfnm{{Geoffrey J. }}\binits{G.}} \AND
  \bauthor{\bsnm{Krishnan},~\bfnm{{Thriyambakam}}\binits{T.}}
(\byear{2008}).
\btitle{The EM algorithm and extensions},
\bedition{2nd} ed.
\bseries{Wiley series in probability and statistics}.
\bpublisher{Wiley}, \baddress{Hoboken, NJ}.
\end{bbook}
\endbibitem

\bibitem[\protect\citeauthoryear{McNeil, Frey and
  Embrechts}{2015}]{mcneil2015quantitative}
\begin{bbook}[author]
\bauthor{\bsnm{McNeil},~\bfnm{Alexander~J}\binits{A.~J.}},
  \bauthor{\bsnm{Frey},~\bfnm{R{\"u}diger}\binits{R.}} \AND
  \bauthor{\bsnm{Embrechts},~\bfnm{Paul}\binits{P.}}
(\byear{2015}).
\btitle{Quantitative Risk Management: Concepts, Techniques and Tools-Revised
  Edition}.
\bpublisher{Princeton University Press}.
\end{bbook}
\endbibitem

\bibitem[\protect\citeauthoryear{Merz and Bl{\"o}schl}{2005}]{mer2005}
\begin{barticle}[author]
\bauthor{\bsnm{Merz},~\bfnm{R.}\binits{R.}} \AND
  \bauthor{\bsnm{Bl{\"o}schl},~\bfnm{G.}\binits{G.}}
(\byear{2005}).
\btitle{Flood frequency regionalisation—spatial proximity vs. catchment
  attributes}.
\bjournal{Journal of Hydrology}
\bvolume{302}
\bpages{283 - 306}.
\end{barticle}
\endbibitem

\bibitem[\protect\citeauthoryear{Mhalla, Chavez-Demoulin and
  Dupuis}{2020}]{mha2020}
\begin{barticle}[author]
\bauthor{\bsnm{Mhalla},~\bfnm{Linda}\binits{L.}},
  \bauthor{\bsnm{Chavez-Demoulin},~\bfnm{Valérie}\binits{V.}} \AND
  \bauthor{\bsnm{Dupuis},~\bfnm{Debbie~J.}\binits{D.~J.}}
(\byear{2020}).
\btitle{Causal mechanism of extreme river discharges in the upper {D}anube
  basin network}.
\bjournal{Journal of the Royal Statistical Society: Series C (Applied
  Statistics)}
\bvolume{69}
\bpages{741-764}.
\end{barticle}
\endbibitem

\bibitem[\protect\citeauthoryear{Mhalla, Hambuckers and
  Lambert}{2020}]{mhalla2020extremal}
\begin{barticle}[author]
\bauthor{\bsnm{Mhalla},~\bfnm{Linda}\binits{L.}},
  \bauthor{\bsnm{Hambuckers},~\bfnm{Julien}\binits{J.}} \AND
  \bauthor{\bsnm{Lambert},~\bfnm{Marie}\binits{M.}}
(\byear{2020}).
\btitle{Extremal connectedness and systemic risk of hedge funds}.
\bjournal{Working Paper (Available at SSRN)}.
\end{barticle}
\endbibitem

\bibitem[\protect\citeauthoryear{Oh and Patton}{2020}]{oh2020dynamic}
\begin{barticle}[author]
\bauthor{\bsnm{Oh},~\bfnm{Dong~Hwan}\binits{D.~H.}} \AND
  \bauthor{\bsnm{Patton},~\bfnm{Andrew~J}\binits{A.~J.}}
(\byear{2020}).
\btitle{Dynamic Factor Copula Models with Estimated Cluster Assignments}.
\bjournal{Working Paper}.
\end{barticle}
\endbibitem

\bibitem[\protect\citeauthoryear{{Committee on the Global Financial
  System}}{2001}]{bis2001}
\begin{btechreport}[author]
\bauthor{\bsnm{{Committee on the Global Financial System}}}
(\byear{2001}).
\btitle{Stress testing by large financial institutions: current practice and
  aggregation issues}
\btype{Technical Report},
\bpublisher{Bank for International Settlements}.
\end{btechreport}
\endbibitem

\bibitem[\protect\citeauthoryear{Overeem, Buishand and
  Holleman}{2008}]{overeem2008}
\begin{barticle}[author]
\bauthor{\bsnm{Overeem},~\bfnm{Aart}\binits{A.}},
  \bauthor{\bsnm{Buishand},~\bfnm{Adri}\binits{A.}} \AND
  \bauthor{\bsnm{Holleman},~\bfnm{Iwan}\binits{I.}}
(\byear{2008}).
\btitle{Rainfall depth-duration-frequency curves and their uncertainties}.
\bjournal{Journal of Hydrology}
\bvolume{348}
\bpages{124--134}.
\end{barticle}
\endbibitem

\bibitem[\protect\citeauthoryear{Pakel, Shephard and
  Sheppard}{2011}]{pakel2011nuisance}
\begin{barticle}[author]
\bauthor{\bsnm{Pakel},~\bfnm{Cavit}\binits{C.}},
  \bauthor{\bsnm{Shephard},~\bfnm{Neil}\binits{N.}} \AND
  \bauthor{\bsnm{Sheppard},~\bfnm{Kevin}\binits{K.}}
(\byear{2011}).
\btitle{Nuisance parameters, composite likelihoods and a panel of {GARCH}
  models}.
\bjournal{Statistica Sinica}
\bvolume{21}
\bpages{307--329}.
\end{barticle}
\endbibitem

\bibitem[\protect\citeauthoryear{Papalexiou et~al.}{2018}]{papalexiou2018}
\begin{barticle}[author]
\bauthor{\bsnm{Papalexiou},~\bfnm{S.~M.}\binits{S.~M.}},
  \bauthor{\bsnm{AghaKouchak},~\bfnm{A.}\binits{A.}},
  \bauthor{\bsnm{Trenberth},~\bfnm{K.~E.}\binits{K.~E.}} \AND
  \bauthor{\bsnm{Foufoula-Georgiou},~\bfnm{E.}\binits{E.}}
(\byear{2018}).
\btitle{Global, Regional, and Megacity Trends in the Highest Temperature of the
  Year: diagnostics and Evidence for Accelerating Trends}.
\bjournal{Earth's Future}
\bvolume{6}
\bpages{71--79}.
\end{barticle}
\endbibitem

\bibitem[\protect\citeauthoryear{Petersen}{2019}]{petersen2019}
\begin{barticle}[author]
\bauthor{\bsnm{Petersen},~\bfnm{L.~K.}\binits{L.~K.}}
(\byear{2019}).
\btitle{Impact of Climate Change on Twenty-First Century Crop Yields in the
  {U.S.}}
\bjournal{Climate}
\bvolume{7}.
\end{barticle}
\endbibitem

\bibitem[\protect\citeauthoryear{Pickands}{1975}]{pic1975}
\begin{barticle}[author]
\bauthor{\bsnm{Pickands},~\bfnm{James}\binits{J.} \bsuffix{III}}
(\byear{1975}).
\btitle{Statistical inference using extreme order statistics}.
\bjournal{Ann. Statist.}
\bvolume{3}
\bpages{119--131}.
\end{barticle}
\endbibitem

\bibitem[\protect\citeauthoryear{Rand}{1971}]{rand1971objective}
\begin{barticle}[author]
\bauthor{\bsnm{Rand},~\bfnm{William~M}\binits{W.~M.}}
(\byear{1971}).
\btitle{Objective criteria for the evaluation of clustering methods}.
\bjournal{Journal of the American Statistical Association}
\bvolume{66}
\bpages{846--850}.
\end{barticle}
\endbibitem

\bibitem[\protect\citeauthoryear{Reich and Shaby}{2019}]{ReichShaby2018}
\begin{barticle}[author]
\bauthor{\bsnm{Reich},~\bfnm{Brian~J.}\binits{B.~J.}} \AND
  \bauthor{\bsnm{Shaby},~\bfnm{Benjamin~A.}\binits{B.~A.}}
(\byear{2019}).
\btitle{A Spatial Markov Model for Climate Extremes}.
\bjournal{Journal of Computational and Graphical Statistics}
\bvolume{28}
\bpages{117--126}.
\bdoi{10.1080/10618600.2018.1482764}
\end{barticle}
\endbibitem

\bibitem[\protect\citeauthoryear{Ribatet}{2019}]{spatialextremes}
\begin{bmanual}[author]
\bauthor{\bsnm{Ribatet},~\bfnm{Mathieu}\binits{M.}}
(\byear{2019}).
\btitle{SpatialExtremes: Modelling Spatial Extremes}
\bnote{R package version 2.0-7.2}.
\end{bmanual}
\endbibitem

\bibitem[\protect\citeauthoryear{Rohrbeck and Tawn}{2021}]{RohrbeckTawn2021}
\begin{barticle}[author]
\bauthor{\bsnm{Rohrbeck},~\bfnm{Christian}\binits{C.}} \AND
  \bauthor{\bsnm{Tawn},~\bfnm{Jonathan~A.}\binits{J.~A.}}
(\byear{2021}).
\btitle{Bayesian Spatial Clustering of Extremal Behavior for Hydrological
  Variables}.
\bjournal{Journal of Computational and Graphical Statistics}
\bvolume{30}
\bpages{91--105}.
\bdoi{10.1080/10618600.2020.1777139}
\end{barticle}
\endbibitem

\bibitem[\protect\citeauthoryear{R{\"o}ttger, Engelke and
  Zwiernik}{2021}]{roe2021}
\begin{bunpublished}[author]
\bauthor{\bsnm{R{\"o}ttger},~\bfnm{Frank}\binits{F.}},
  \bauthor{\bsnm{Engelke},~\bfnm{Sebastian}\binits{S.}} \AND
  \bauthor{\bsnm{Zwiernik},~\bfnm{Piotr}\binits{P.}}
(\byear{2021}).
\btitle{Total positivity in multivariate extremes}.
\bnote{Available from \texttt{https://arxiv.org/abs/2112.14727}}.
\end{bunpublished}
\endbibitem

\bibitem[\protect\citeauthoryear{Sadok and Krishna~Jagadish}{2020}]{sadok2020}
\begin{barticle}[author]
\bauthor{\bsnm{Sadok},~\bfnm{W.}\binits{W.}} \AND
  \bauthor{\bsnm{Krishna~Jagadish},~\bfnm{S.~V.}\binits{S.~V.}}
(\byear{2020}).
\btitle{The Hidden Costs of Nighttime Warming on Yields}.
\bjournal{Trends in Plant Science}
\bvolume{25}
\bpages{644--651}.
\end{barticle}
\endbibitem

\bibitem[\protect\citeauthoryear{Segal, Shaliastovich and
  Yaron}{2015}]{segal2015good}
\begin{barticle}[author]
\bauthor{\bsnm{Segal},~\bfnm{Gill}\binits{G.}},
  \bauthor{\bsnm{Shaliastovich},~\bfnm{Ivan}\binits{I.}} \AND
  \bauthor{\bsnm{Yaron},~\bfnm{Amir}\binits{A.}}
(\byear{2015}).
\btitle{Good and bad uncertainty: macroeconomic and financial market
  implications}.
\bjournal{Journal of Financial Economics}
\bvolume{117}
\bpages{369--397}.
\end{barticle}
\endbibitem

\bibitem[\protect\citeauthoryear{Smith}{1985}]{smi1985}
\begin{barticle}[author]
\bauthor{\bsnm{Smith},~\bfnm{R.~L.}\binits{R.~L.}}
(\byear{1985}).
\btitle{Maximum likelihood estimation in a class of nonregular cases.}
\bjournal{Biometrika}
\bvolume{72}
\bpages{67-90}.
\end{barticle}
\endbibitem

\bibitem[\protect\citeauthoryear{Su, Shi and
  Phillips}{2016}]{su2016identifying}
\begin{barticle}[author]
\bauthor{\bsnm{Su},~\bfnm{Liangjun}\binits{L.}},
  \bauthor{\bsnm{Shi},~\bfnm{Zhentao}\binits{Z.}} \AND
  \bauthor{\bsnm{Phillips},~\bfnm{Peter~CB}\binits{P.~C.}}
(\byear{2016}).
\btitle{Identifying latent structures in panel data}.
\bjournal{Econometrica}
\bvolume{84}
\bpages{2215--2264}.
\end{barticle}
\endbibitem

\bibitem[\protect\citeauthoryear{Velthoen et~al.}{2021}]{vel2021}
\begin{bmisc}[author]
\bauthor{\bsnm{Velthoen},~\bfnm{Jasper}\binits{J.}},
  \bauthor{\bsnm{Dombry},~\bfnm{Clément}\binits{C.}},
  \bauthor{\bsnm{Cai},~\bfnm{Juan-Juan}\binits{J.-J.}} \AND
  \bauthor{\bsnm{Engelke},~\bfnm{Sebastian}\binits{S.}}
(\byear{2021}).
\btitle{Gradient boosting for extreme quantile regression}.
\end{bmisc}
\endbibitem

\bibitem[\protect\citeauthoryear{Vignotto, Engelke and
  Zscheischler}{2021}]{vig2021}
\begin{barticle}[author]
\bauthor{\bsnm{Vignotto},~\bfnm{Edoardo}\binits{E.}},
  \bauthor{\bsnm{Engelke},~\bfnm{Sebastian}\binits{S.}} \AND
  \bauthor{\bsnm{Zscheischler},~\bfnm{Jakob}\binits{J.}}
(\byear{2021}).
\btitle{Clustering bivariate dependencies of compound precipitation and wind
  extremes over Great Britain and Ireland}.
\bjournal{Weather and Climate Extremes}
\bvolume{32}.
\bdoi{https://doi.org/10.1016/j.wace.2021.100318}
\end{barticle}
\endbibitem

\bibitem[\protect\citeauthoryear{Wang and Su}{2021}]{wang2021identifying}
\begin{barticle}[author]
\bauthor{\bsnm{Wang},~\bfnm{Wuyi}\binits{W.}} \AND
  \bauthor{\bsnm{Su},~\bfnm{Liangjun}\binits{L.}}
(\byear{2021}).
\btitle{Identifying latent group structures in nonlinear panels}.
\bjournal{Journal of Econometrics}
\bvolume{220}
\bpages{272--295}.
\end{barticle}
\endbibitem

\bibitem[\protect\citeauthoryear{Wang et~al.}{2016}]{wang2016gev}
\begin{barticle}[author]
\bauthor{\bsnm{Wang},~\bfnm{Jiali}\binits{J.}},
  \bauthor{\bsnm{Han},~\bfnm{Yuefeng}\binits{Y.}},
  \bauthor{\bsnm{Stein},~\bfnm{Michael~L.}\binits{M.~L.}},
  \bauthor{\bsnm{Kotamarthi},~\bfnm{Veerabhadra~R.}\binits{V.~R.}} \AND
  \bauthor{\bsnm{Huang},~\bfnm{Whitney~K.}\binits{W.~K.}}
(\byear{2016}).
\btitle{Evaluation of dynamically downscaled extreme temperature using a
  spatially-aggregated generalized extreme value (GEV) model}.
\bjournal{Climate Dynamics}
\bvolume{47}
\bpages{2833--2848}.
\end{barticle}
\endbibitem

\bibitem[\protect\citeauthoryear{White}{1994}]{white1996estimation}
\begin{bbook}[author]
\bauthor{\bsnm{White},~\bfnm{Halbert}\binits{H.}}
(\byear{1994}).
\btitle{Estimation, Inference and Specification Analysis}.
\bseries{Econometric Society Monographs No. 22}.
\bpublisher{Cambridge University Press}.
\end{bbook}
\endbibitem

\bibitem[\protect\citeauthoryear{Wooldridge}{1994}]{wooldridge1994estimation}
\begin{barticle}[author]
\bauthor{\bsnm{Wooldridge},~\bfnm{Jeffrey~M}\binits{J.~M.}}
(\byear{1994}).
\btitle{Estimation and inference for dependent processes}.
\bjournal{Handbook of Econometrics}
\bvolume{4}
\bpages{2639--2738}.
\end{barticle}
\endbibitem

\bibitem[\protect\citeauthoryear{Zhao, Zhang and Chen}{2018}]{zhao2018modeling}
\begin{barticle}[author]
\bauthor{\bsnm{Zhao},~\bfnm{Zifeng}\binits{Z.}},
  \bauthor{\bsnm{Zhang},~\bfnm{Zhengjun}\binits{Z.}} \AND
  \bauthor{\bsnm{Chen},~\bfnm{Rong}\binits{R.}}
(\byear{2018}).
\btitle{Modeling maxima with autoregressive conditional {F}r{\'e}chet model}.
\bjournal{Journal of Econometrics}
\bvolume{207}
\bpages{325--351}.
\end{barticle}
\endbibitem

\bibitem[\protect\citeauthoryear{Zwiers and Kharin}{1998}]{zwiers1998}
\begin{barticle}[author]
\bauthor{\bsnm{Zwiers},~\bfnm{F.~W.}\binits{F.~W.}} \AND
  \bauthor{\bsnm{Kharin},~\bfnm{V.~V.}\binits{V.~V.}}
(\byear{1998}).
\btitle{Changes in the Extremes of the Climate Simulated by {CCC} {GCM2} under
  {CO2} Doubling}.
\bjournal{Journal of Climate}
\bvolume{11}
\bpages{2200--2222}.
\end{barticle}
\endbibitem

\end{thebibliography}

\begin{acks}[Acknowledgments]
\red{
The authors wish to thank the Associate Editor
and two anonymous referees for helpful comments that improved the paper.}
\end{acks}

\begin{funding}
\red{
The first author was supported by the Natural Sciences and Engineering Research
Council of Canada RGPIN-2016-04114 and the HEC Foundation. The second author was supported by the Swiss National Science Foundation (Grant 186858). The third author was supported by a Modigliani Research Grant of the UniCredit Foundation for the project ``On the determinants of time-varying tail risk''.}
\end{funding}

\setcounter{section}{0}
\setcounter{subsection}{0}
\setcounter{equation}{0}
\renewcommand{\thesection}{A.\arabic{section}}
\renewcommand{\thesubsection}{A.\arabic{subsection}}
\renewcommand{\theequation}{A.\arabic{equation}}

\renewcommand\theHsubsection{A.\thesubsection}
\renewcommand\theHequation{A.\theequation}

\section*{Appendix}

\section{Consistency of EM algorithm}
\label{app:consistency}
	\noindent Let $\left(\mathbf{Y}_t, \mathbf{X}_t\right)$ be an $N\times(K+1)$ matrix of random variables. Let $\boldsymbol{\theta} \in \boldsymbol{\Theta}$ be a $P$-dimensional vector of parameters and let $\boldsymbol{\tau} \in \mathcal{G}$ be a vector of $N$ integers denoting the assignment of each individual to one of $G$ groups.
	
	We reformulate the iteration of the EM algorithm for the joint estimation of the model parameters and group assignments in Section~\ref{Sec:GroupPanel} in terms of M-estimators. At the $(j+1)$th iteration, we have,
	\begin{enumerate}
	\item[1.1]  \textbf{Maximization}.
	$
	\widehat{\boldsymbol{\theta}}^{(j+1)} = \arg\min_{\boldsymbol{\theta}} 	S_T\left(\boldsymbol{\theta}, \widehat{\boldsymbol{\tau}}^{(j)}\right)
	$
	\item[1.2]  \textbf{Expectation}.
	$
	\widehat{\tau}^{(j+1)}_i = \arg \min_{g \in \left\{1,\dots,G\right\}} S_T\left(\widehat{\boldsymbol{\theta}}^{(j+1)}, \widehat{\boldsymbol{\tau}}^{(j)}_{i,g}\right)
	$
	\end{enumerate}
	where $\widehat{\boldsymbol{\tau}}^{(j)}_{i,g}$ is equal to $\widehat{\boldsymbol{\tau}}^{(j)}$ except that the $i$th element is equal to $g$, and 
	\[
	S_T(\boldsymbol{\theta}, \boldsymbol{\tau})  = \frac{1}{T}\sum_{t=1}^{T} \rho \left(\mathbf{Y}_t, \mathbf{X}_t, \boldsymbol{\theta}, \boldsymbol{\tau} \right)
	\]
	is the objective function used in the M-estimation, with $\rho \left(\mathbf{Y}_t, \mathbf{X}_t, \boldsymbol{\theta}, \boldsymbol{\tau} \right)$ some function of $(\mathbf{Y}_t, \mathbf{X}_t)$, the parameters $\boldsymbol{\theta}\in\boldsymbol{\Theta}$ and $\boldsymbol{\tau} \in \mathcal{G}$. The EM estimates $(\widehat{\boldsymbol{\theta}}_T, \widehat{\boldsymbol{\tau}}_T)$ are obtained by iterating these two steps until convergence  of the EM algorithm. In Section \ref{Sec:GroupPanel}, we define the criterion function $\rho \left(\mathbf{Y}_t, \mathbf{X}_t, \boldsymbol{\theta}, \boldsymbol{\tau} \right)$ as
	\[
	\rho \left(\mathbf{Y}_t, \mathbf{X}_t, \boldsymbol{\theta}, \boldsymbol{\tau} \right) = - \sum_{i=1}^N \log h\left(Y_{it}\mid X_{it},\boldsymbol{\theta}, \boldsymbol{\tau}\right)
	\]
	where $\log h$ is the log-likelihood of the GEV distribution and $X_{it}$, $\boldsymbol{\theta}$, and $\boldsymbol{\tau}$ enter the GEV parameters $\mu$, $\sigma$, and $\xi$ as in \eqref{grouped_panel_parms}. 
	
	We provide conditions under which $\left(\widehat{\boldsymbol{\theta}}_T, \widehat{\boldsymbol{\tau}}_T\right)$ are consistent for their true counterparts $\left(\boldsymbol{\theta}_0, \boldsymbol{\tau}_0\right)$, where
	\[
	\boldsymbol{\theta}_0 = \arg \min_{\boldsymbol{\theta}} S(\boldsymbol{\theta}, \boldsymbol{\tau}_0)
	\]	
	\[
	S(\boldsymbol{\theta}, \boldsymbol{\tau}) = \mathbb{E} \left[ S_T(\boldsymbol{\theta}, \boldsymbol{\tau}) \right] = \mathbb{E} \left[ \rho \left(\mathbf{Y}_t, \mathbf{X}_t, \boldsymbol{\theta}, \boldsymbol{\tau} \right) \right].
	\]
	Note that labels attached to the groups are arbitrary and there exists a set of correct group assignments, that we denote $\mathcal{G}_0$. We can thus consistently estimate $\boldsymbol{\tau}_0$ up to re-labeling of the groups.

	\vspace*{0.2in}
	\noindent \textbf{Assumption 1:} $\left\{ \left(\mathbf{Y}_t, \mathbf{X}_t\right) : t=1,2,\dots \right\}$ is a stationary ergodic sequence.
	
	\noindent \textbf{Assumption 2:} $\boldsymbol{\Theta}$ is compact.
	
	\noindent \textbf{Assumption 3:} For each $\boldsymbol{\tau}\in\mathcal{G}$ and  $\left(\mathbf{Y}_t, \mathbf{X}_t\right)$, $\rho\left(\mathbf{Y}_t, \mathbf{X}_t, \boldsymbol{\theta}, \boldsymbol{\tau} \right)$ is continuous in $\boldsymbol{\Theta}$. For each $\boldsymbol{\tau} \in \mathcal{G}$ and for all $\boldsymbol{\theta} \in \boldsymbol{\Theta}$, $\rho\left(\mathbf{Y}_t, \mathbf{X}_t, \boldsymbol{\theta}, \boldsymbol{\tau} \right)$ is measurable.
	
	\noindent \textbf{Assumption 4:} For each $\boldsymbol{\tau} \in \mathcal{G}$ and for all $\boldsymbol{\theta} \in \boldsymbol{\Theta}$, it holds that:
	\begin{itemize}
		\item[(i)] $\mathbb{E}\left[\left|\rho \left(\mathbf{Y}_t, \mathbf{X}_t, \boldsymbol{\theta}, \boldsymbol{\tau} \right) \right| \right] < \infty$;
		\item[(ii)] $\mathbb{E}\left[\left\| \nabla_{\boldsymbol{\theta}} \rho \left(\mathbf{Y}_t, \mathbf{X}_t, \boldsymbol{\theta}, \boldsymbol{\tau} \right) \right\|_1 \right]< \infty$;
		\item[(iii)] $\mathbb{E}\left[\left\| \nabla_{\boldsymbol{\theta}\boldsymbol{\theta}} \rho \left(\mathbf{Y}_t, \mathbf{X}_t, \boldsymbol{\theta}, \boldsymbol{\tau} \right) \right\|_1\right] < \infty$  .
	\end{itemize} 
	
	\noindent \textbf{Assumption 5}: For any $\boldsymbol{\tau} \in \mathcal G$, let $\eta_{\boldsymbol{\tau}}(\epsilon)$ be an $\epsilon$-neighborhood of  $\boldsymbol{\theta}^*(\boldsymbol{\tau})$. For all $\epsilon>0$,
	\begin{itemize}
		\item[(i)]  $
		\inf_{\boldsymbol{\theta} \in \left\{\boldsymbol{\Theta}\backslash \eta_{\boldsymbol{\tau}}(\epsilon) \right\}} S(\boldsymbol{\theta}, \boldsymbol{\tau})>S(\boldsymbol{\theta}^* (\boldsymbol{\tau}), \boldsymbol{\tau}), \quad \forall \boldsymbol{\tau}\in\mathcal{G}.
		$
	\end{itemize}
	Moreover, for $\boldsymbol{\tau}=\boldsymbol{\tau}_0$ we have
	\begin{itemize}	
		\item[(ii)] $
		\inf_{\left(\boldsymbol{\theta}, \boldsymbol{\tau}\right) \in \boldsymbol{\Theta} \times \left\{\mathcal{G}\backslash \mathcal{G}_0 \right\}} S(\boldsymbol{\theta}, \boldsymbol{\tau})>S(\boldsymbol{\theta}_0, \boldsymbol{\tau}_0)
		$
	\end{itemize}
        
	\vspace*{0.2in} 
	
	\begin{theorem}\label{thm1}
        Under Assumptions 1--5, we have that $\widehat{\boldsymbol{\theta}}_T \xrightarrow{p} \boldsymbol{\theta}_0$ and $\widehat{\boldsymbol{\tau}}_T \xrightarrow{p} \boldsymbol{\tau}_0$.
	\end{theorem}
	\noindent \textbf{Proof}. Define the feasible and unfeasible profile estimators
		\[
	\widetilde{\boldsymbol{\theta}}_T (\boldsymbol{\tau}) =  \arg \min_{\boldsymbol{\theta}} \frac{1}{T} \sum_{t=1}^{T} \rho \left(\mathbf{Y}_t, \mathbf{X}_t, \boldsymbol{\theta}, \boldsymbol{\tau} \right)
	\]
	\[
	\boldsymbol{\theta}^*(\boldsymbol{\tau}) = \arg \min_{\boldsymbol{\theta}} S(\boldsymbol{\theta}, \boldsymbol{\tau}).
	\]
        We first show that $\widetilde{\boldsymbol{\theta}}_T (\boldsymbol{\tau}) \xrightarrow{p} \boldsymbol{\theta}^* (\boldsymbol{\tau})$ for a given $\boldsymbol{\tau}$. Theorem 3.4 of \cite{white1996estimation} implies this result if we can show that: (i) $\boldsymbol{\Theta}$ is compact; (ii) $\rho\left(\mathbf{Y}_t, \mathbf{X}_t, \boldsymbol{\theta}, \boldsymbol{\tau} \right)$ is measurable and conti\-nuous on $\boldsymbol{\Theta}$; (iii) $\boldsymbol{\theta}^* (\boldsymbol{\tau})$ is the unique minimizer of $S(\boldsymbol{\theta},\boldsymbol{\tau})$ on $\boldsymbol{\Theta}$; (iv) $S_T(\boldsymbol{\theta},\boldsymbol{\tau})$ converges to $S(\boldsymbol{\theta},\boldsymbol{\tau})$ uniformly on $\boldsymbol{\Theta}$. Conditions (i)-(ii) are satisfied by Assumptions 2 and 3. Condition (iii) is satisfied by Assumption 5(i). To show that condition (iv) is satisfied we verify Theorem 4.1 of \cite{wooldridge1994estimation}: Our Assumption 1 satisfies their stationary and ergodic condition; our Assumptions 2 and 3 match their Assumptions (i)-(ii); our Assumption 4(i) matches their Assumption (iii).
	
	As $\widetilde{\boldsymbol{\theta}}_T (\boldsymbol{\tau}) \xrightarrow{p} \boldsymbol{\theta}^* (\boldsymbol{\tau})$ for each $\boldsymbol{\tau} \in \mathcal{G}$, $\widehat{\boldsymbol{\tau}}_T \xrightarrow{p} \boldsymbol{\tau}_0$ implies $\widehat{\boldsymbol{\theta}}_T \xrightarrow{p} \boldsymbol{\theta}_0$, see also Theorem 4.3 of \cite{wooldridge1994estimation}. To prove consistency of $\widehat{\boldsymbol{\tau}}_T $ to $\boldsymbol{\tau}_0$ we follow \cite{oh2020dynamic} showing that (i) $\boldsymbol{\tau}_0$ uniquely minimizes $S(\boldsymbol{\theta}^*(\boldsymbol{\tau}),\boldsymbol{\tau})$ and (ii) S$_T(\widetilde{\boldsymbol{\theta}}_T(\boldsymbol{\tau}),\boldsymbol{\tau})$ converges pointwise to $S(\boldsymbol{\theta}^*(\boldsymbol{\tau}),\boldsymbol{\tau})$ for each $\boldsymbol{\tau} \in \mathcal{G}$.  Condition (i) is satisfied by Assumption 5(ii), so we focus on showing condition (ii). We have 
	\[
	S_T(\widetilde{\boldsymbol{\theta}}_T(\boldsymbol{\tau}),\boldsymbol{\tau})-S(\boldsymbol{\theta}^*(\boldsymbol{\tau}),\boldsymbol{\tau}) = S_T(\widetilde{\boldsymbol{\theta}}_T(\boldsymbol{\tau}),\boldsymbol{\tau}) - S_T(\boldsymbol{\theta}^*(\boldsymbol{\tau}),\boldsymbol{\tau}) + S_T(\boldsymbol{\theta}^*(\boldsymbol{\tau}),\boldsymbol{\tau}) - S(\boldsymbol{\theta}^*(\boldsymbol{\tau}),\boldsymbol{\tau})
	\]
        Under Assumptions 1 and 4(i), the weak law of large numbers guarantees that $S_T(\boldsymbol{\theta}^*(\boldsymbol{\tau}),\boldsymbol{\tau}) - S(\boldsymbol{\theta}^*(\boldsymbol{\tau}),\boldsymbol{\tau}) = o_p(1)$. Further,
	\[
	S_T(\widetilde{\boldsymbol{\theta}}_T(\boldsymbol{\tau}),\boldsymbol{\tau}) - S_T(\boldsymbol{\theta}^*(\boldsymbol{\tau}),\boldsymbol{\tau})  =  \frac{1}{T} \sum_{t=1}^{T}  \left[ \rho \left(\mathbf{Y}_t, \mathbf{X}_t, \widetilde{\boldsymbol{\theta}}_T (\boldsymbol{\tau}) , \boldsymbol{\tau} \right) - \rho \left(\mathbf{Y}_t, \mathbf{X}_t, \boldsymbol{\theta}^*(\boldsymbol{\tau}), \boldsymbol{\tau} \right)\right]
	\]
	From a mean value expansion of $\rho \left(\mathbf{Y}_t, \mathbf{X}_t, \widetilde{\boldsymbol{\theta}}_T (\boldsymbol{\tau}) , \boldsymbol{\tau} \right)$ around $\boldsymbol{\theta}^*(\boldsymbol{\tau})$ we obtain
	\[
	\small
	\begin{array}{rl}
		S_T(\widetilde{\boldsymbol{\theta}}_T(\boldsymbol{\tau}),\boldsymbol{\tau}) - S_T(\boldsymbol{\theta}^*(\boldsymbol{\tau}),\boldsymbol{\tau}) &= \left(\frac{1}{T} \sum_{t=1}^{T} \nabla_{\boldsymbol{\theta}} \rho \left(\mathbf{Y}_t, \mathbf{X}_t, \boldsymbol{\theta}^*(\boldsymbol{\tau}), \boldsymbol{\tau} \right) \right)' \left(  \widetilde{\boldsymbol{\theta}}(\boldsymbol{\tau})-\boldsymbol{\theta}^*(\boldsymbol{\tau})\right) 	 \\
		& + \frac{1}{2} \left(  \widetilde{\boldsymbol{\theta}}(\boldsymbol{\tau})-\boldsymbol{\theta}^*(\boldsymbol{\tau})\right)' \left(\frac{1}{T} \sum_{t=1}^{T}\nabla_{\boldsymbol{\theta}\boldsymbol{\theta}} \rho \left(\mathbf{Y}_t, \mathbf{X}_t, \dot{\boldsymbol{\theta}}(\boldsymbol{\tau}), \boldsymbol{\tau} \right) \right) \left(  \widetilde{\boldsymbol{\theta}}(\boldsymbol{\tau})-\boldsymbol{\theta}^*(\boldsymbol{\tau})\right)
	\end{array}
	\]
	where $\dot{\boldsymbol{\theta}}(\boldsymbol{\tau}) = \lambda \widetilde{\boldsymbol{\theta}}(\boldsymbol{\tau}) + (1-\lambda) \boldsymbol{\theta}^*(\boldsymbol{\tau})$ for some $\lambda\in [0,1]$.
	Assumptions 1, 5(ii), 5(iii) guarantee that the gradient and hessian terms converge in probability to a finite limit by the weak law of large numbers. Since $\widetilde{\boldsymbol{\theta}}(\boldsymbol{\tau})\xrightarrow{p}\boldsymbol{\theta}^*(\boldsymbol{\tau})$, we have $	S_T(\widetilde{\boldsymbol{\theta}}_T(\boldsymbol{\tau}),\boldsymbol{\tau}) - S_T(\boldsymbol{\theta}^*(\boldsymbol{\tau}),\boldsymbol{\tau})=o_p(1)$.

\section{A grouped panel GP  regression model}

       We outline here an approach to extend the methodology of this paper to grouped panels of generalized Pareto (GP) distributions. Let $\left(Z_{i,t},\mathbf{X}_{i,t} \right)$ be the observations for the $i$th individual at time $t$,  with $i=1,\dots,N$ and $t=1,\dots,T$. Here, the $Z_{i,t}$ can be seen as daily observations as opposed to the $Y_{i,t}$ that represented block maxima over a certain block size (e.g., a year).
       For an intermediate proba\-bility level $p_0$, the GP distribution~\citep{pic1975} is used to model the exceendances $(Z_{i,t} \mid Z_{i,t} > Q^{p_0}_{i,t},\mathbf{X}_{i,t}) $ over its conditional intermediate quantile $Q^{p_0}_{i,t}$ at level $p_0$ given the covariate vector $\mathbf{X}_{i,t}$. More precisely, the distribution function of these exceedances is approximated by a GP distribution with covariate dependent parameters
 given by
\begin{equation*}
  W(z\mid \sigma_{i,t}, \xi_{i,t}) = 1-\left(1 + \frac{\xi_{i,t}}{\sigma_{i,t}} z \right)_+^{-1/\xi_{i,t}},\quad z > 0,
\end{equation*}
where $\sigma_{i,t}>0$ and $\xi_{i,t}\in \mathbb R$ are the conditional scale and shape parameters, respectively. To model a grouped panel GP regression, similarly to~\eqref{grouped_panel}, we may assume that these parameters satisfy
\begin{align}
\label{grouped_GP_parms}
\begin{array}{rll}
\sigma_{i,t}(\tau_i) &=& {e_\sigma} \left(\boldsymbol{\gamma}_{(\tau_i)}^{\top} \mathbf{X}_{i,t}\right) \\
\xi_{i,t}(\tau_i) &=& {e_\xi} \left(\boldsymbol{\delta}_{(\tau_i)}^{\top} \mathbf{X}_{i,t} \right)
\end{array}
\end{align}
where $\tau_i \in \{1,\dots,G\}$ denotes the group membership of the $i$th individual
and $\boldsymbol{\theta}_{(g)}=\left(\boldsymbol{\gamma}_{(g)},\boldsymbol{\delta}_{(g)}\right) \in \boldsymbol{\Theta} \subset \mathbb{R}^P$ is the vector of regression parameters for the $g$th group, $g \in \{1,\dots,G\}$.

We can then employ an EM algorithm, similar to the one of the grouped panel GEV model in Section~\ref{Sec:GroupPanel}, that is based on the exceedances of the data set  $\left(Z_{i,t},\mathbf{X}_{i,t} \right)$, $i=1,\dots,N$, $t=1,\dots,T$, over their respective $p_0$ quantiles. That is, we would replace the GEV log-likelihoods in the EM algorithm by the log-likelihood of the GP distribution given by
$$ \log w(z \mid \sigma_{i,t}, \xi_{i,t}) =  -\log \sigma_{i,t} - \left(1 + \frac{1}{\xi_{i,t}}\right)\log \left\{1 + \frac{\xi_{i,t}}{\sigma_{i,t}} (Z_{i,t} - Q^{p_0}_{i,t})\right\}.$$
A difficulty in this panel GP model is that it relies on a two-step procedure. First one needs a good estimation of the intermediate conditional quantiles $Q^{p_0}_{i,t}$ to define the exceedances that are provided to the EM algorithm. This can either be done by an additional panel quantile regression \citep{gu2019panel}, or with any other quantile regression method that the modeler considers suitable.

\end{document}